\documentstyle[psfig]{mn}

\newif\ifAMStwofonts
\def\lsim{\raise0.3ex\hbox{$<$}\kern-0.75em{\lower0.65ex\hbox{$\sim$}}}
\def\gsim{\raise0.3ex\hbox{$>$}\kern-0.75em{\lower0.65ex\hbox{$\sim$}}}

\ifoldfss
  \ifCUPmtlplainloaded \else
    \NewTextAlphabet{textbfit} {cmbxti10} {}
    \NewTextAlphabet{textbfss} {cmssbx10} {}
    \NewMathAlphabet{mathbfit} {cmbxti10} {} % for math mode
    \NewMathAlphabet{mathbfss} {cmssbx10} {} %  "   "    "
  \fi
  \ifAMStwofonts
    \ifCUPmtlplainloaded \else
      \NewSymbolFont{upmath} {eurm10}
      \NewSymbolFont{AMSa} {msam10}
      \NewMathSymbol{\upi}     {0}{upmath}{19}
      \NewMathSymbol{\umu}     {0}{upmath}{16}
      \NewMathSymbol{\upartial}{0}{upmath}{40}
      \NewMathSymbol{\leqslant}{3}{AMSa}{36}
      \NewMathSymbol{\geqslant}{3}{AMSa}{3E}

       \let\le=\leqslant
       
    \fi
  \fi
\fi % End of OFSS

\ifnfssone
  \newmathalphabet{\mathit}
  \addtoversion{normal}{\mathit}{cmr}{m}{it}
  \addtoversion{bold}{\mathit}{cmr}{bx}{it}
  \newmathalphabet{\mathbfit} % math mode version of \textbfit{..}
  \addtoversion{normal}{\mathbfit}{cmr}{bx}{it}
  \addtoversion{bold}{\mathbfit}{cmr}{bx}{it}
  \newmathalphabet{\mathbfss} % math mode version of \textbfss{..}
  \addtoversion{normal}{\mathbfss}{cmss}{bx}{n}
  \addtoversion{bold}{\mathbfss}{cmss}{bx}{n}
  \ifAMStwofonts
    \ifCUPmtlplainloaded \else
      \UseAMStwoboldmath
      \makeatletter
      \new@mathgroup\upmath@group
      \define@mathgroup\mv@normal\upmath@group{eur}{m}{n}
      \define@mathgroup\mv@bold\upmath@group{eur}{b}{n}
      \edef\UPM{\hexnumber\upmath@group}
      \new@mathgroup\amsa@group
      \define@mathgroup\mv@normal\amsa@group{msa}{m}{n}
      \define@mathgroup\mv@bold\amsa@group{msa}{m}{n}
      \edef\AMSa{\hexnumber\amsa@group}
      \makeatother
      \mathchardef\upi="0\UPM19
      \mathchardef\umu="0\UPM16
      \mathchardef\upartial="0\UPM40
      \mathchardef\leqslant="3\AMSa36
      \mathchardef\geqslant="3\AMSa3E

       \let\le=\leqslant

    \fi
  \fi
\fi % End of NFSS release 1

\ifnfsstwo
  \DeclareMathAlphabet{\mathbfit}{OT1}{cmr}{bx}{it}
  \SetMathAlphabet\mathbfit{bold}{OT1}{cmr}{bx}{it}
  \DeclareMathAlphabet{\mathbfss}{OT1}{cmss}{bx}{n}
  \SetMathAlphabet\mathbfss{bold}{OT1}{cmss}{bx}{n}
  \ifAMStwofonts
    \ifCUPmtlplainloaded \else
      \DeclareSymbolFont{UPM}{U}{eur}{m}{n}
      \SetSymbolFont{UPM}{bold}{U}{eur}{b}{n}
      \DeclareSymbolFont{AMSa}{U}{msa}{m}{n}
      \DeclareMathSymbol{\upi}{0}{UPM}{"19}
      \DeclareMathSymbol{\umu}{0}{UPM}{"16}
      \DeclareMathSymbol{\upartial}{0}{UPM}{"40}
      \DeclareMathSymbol{\leqslant}{3}{AMSa}{"36}
      \DeclareMathSymbol{\geqslant}{3}{AMSa}{"3E}

       \let\le=\leqslant

    \fi
  \fi
\fi % End of NFSS release 2

\ifCUPmtlplainloaded \else
  \ifAMStwofonts \else % If no AMS fonts
    \def\upi{\pi}
    \def\umu{\mu}
    \def\upartial{\partial}
  \fi
\fi

\title[The Galaxy Population of Cl1601+42]{The Galaxy Population of Cl1601+42 at $z$~=~0.54}
\author[Dahl\'{e}n, Fransson \& N\"{a}slund]{Tomas Dahl\'{e}n\thanks{E-mail: tomas@astro.su.se}, Claes Fransson \& Magnus N\"{a}slund \\ Stockholm Observatory, Department of Astronomy, SCFAB, SE--106 91 Stockholm, Sweden}
\date{Accepted .
      Received ;
      }
\pagerange{\pageref{firstpage}--\pageref{lastpage}}
\pubyear{2001}

\begin{document}

\maketitle

\label{firstpage}

\begin{abstract}
Photometric redshifts are used to determine the rest frame luminosity function (LF) of both early-type and late-type galaxies to $M_B \sim$ -17.6 for  the cluster Cl1601+42 at $z$ = 0.54. The total LF shows a steep faint-end slope $\alpha \sim$ -1.4, indicating the existence of a numerous population of dwarf galaxies. Luminous galaxies, with $M_B~\lsim~-19.5$ are mostly red, early-type galaxies, with a LF best described by a Gaussian. Faint galaxies are predominantly blue, late-type galaxies, well fitted by a Schechter function with $\alpha \sim$ -1.7. Compared to clusters at lower redshift, the steepening of the faint end starts at brighter magnitudes for Cl1601+42, which may indicate a brightening of todays dwarf population relative to the giant population with increasing redshift. 
Early-type galaxies are centrally concentrated, and dominate the core region, implying that the radial gradient of early-type galaxies seen in local clusters is already established at $z~\sim~0.5$. Bright, late-type galaxies are rare, consistent with a decrease in star formation in field galaxies as they are accreted on to the cluster, while faint, blue galaxies are evenly distributed across the cluster, except for a depletion in the core region.
The blue fraction is $f_B \sim$ 0.15, which is somewhat lower than the Butcher-Oemler average at $z~\sim~0.5$. The value of $f_B$ is found to increase with limiting magnitude and with radius from the centre.
\end{abstract}
\section{Introduction}
Butcher \& Oemler (1978a; 1984, hereafter BO84) found that the fraction of blue galaxies increases dramatically between redshifts $z$ = 0 and $z\sim$ 0.5, now usually referred to as the Butcher--Oemler effect (hereafter BO effect). This trend has been confirmed by e.g. Rakos \& Schombert \shortcite{ra95}, Ellingson et al. \shortcite{el01}, Kodama \& Bower \shortcite{ko01} and Margoniner et al. \shortcite{ma01}. The increase of blue galaxies is explained by a higher rate of star formation in cluster galaxies at earlier epochs. The cause of this higher star formation rate is, however, not fully understood. Butcher \& Oemler (1978a) noted that the excess of blue galaxies at higher redshifts compared to todays clusters approximately matches the excess of S0 galaxies in local clusters. If such a change in the morphological mix is caused by an evolution of the blue galaxies into S0 galaxies, once the gas in the star forming blue galaxies is exhausted or tidally stripped, this could lead to the observed BO effect. The evolution of the fraction of S0's between high and low redshift clusters is discussed in e.g. Dressler et al. \shortcite{dr97} and Fasano et al. \shortcite{fa00}.

Bower \shortcite{bo91} and Kauffmann \shortcite{ka95} explain the BO effect as a natural consequence of the hierarchical clustering model for structure evolution in the Universe. In this picture structures, such as clusters, are built up from bottom up, via repeated mergers of subunits. Bower shows that the infall rate of field galaxies to clusters increases with look-back time, leading to a higher fraction of blue galaxies in the clusters. The trend found is not as strong as in the BO84 observations, but can be fitted if the blue fraction in the field increases with redshift. 

Kauffmann shows that high redshift clusters assemble during a much shorter time interval than comparable clusters today. This in turn implies a higher rate of infall of galaxies from the field at higher redshift, producing the BO trend. However, the strong increase in the blue fraction of the observed clusters is explained only in high-density ($\Omega_M~=~1$) CDM models, or MDM models. Low-density CDM models have difficulty in explaining the strong trend at $z~<~0.4$ \cite{ka95}. Recent simulations by Diaferio et al. \shortcite{di01} find that a BO trend is found also in a low-density model ($\Omega_M~=0.3, \Omega_{\Lambda}~=~0.7$), but significantly weaker than in observations.  

It is not clear if star formation is triggered in the infalling galaxies, possibly creating a burst, or if star formation in the field galaxies is simply quenched over some period of time. Both these scenarios seem compatible with the observed BO effect.

The results from BO84 also show that open clusters have a larger fraction of blue galaxies than the more centrally concentrated, compact clusters. (A formal definition of open and compact clusters is given in Section 4.1.) This can be explained in the context of the hierarchical scenario, if the smaller concentration in open clusters is a reflection of a more recent assembly of its members. 

The presence of substructure in a cluster can be a sign of recent subcluster-subcluster merger. From a study of three clusters with signs of substructure at $0.1<z<0.2$, Metevier, Romer \& Ulmer \shortcite{me00} find that at least two of these have an exceptionally large fraction of blue galaxies. This supports a hierarchical scenario where mergers induce star formation in the cluster galaxies.

Furthermore, results from Margoniner et al. \shortcite{ma01} indicate that poor clusters in general have higher blue fractions than rich clusters at comparable redshifts. 
It is therefore clear that the BO effect is not a simple redshift effect, but must be viewed as a consequence of several factors. These include richness, central concentration, epoch of formation and presence of substructure.

A further prediction of the hierarchical CDM model is that the number density of galaxies should be dominated by low mass systems. This is reflected in the luminosity function (hereafter LF), defined as the number density of galaxies per unit luminosity, or per unit magnitude. For the hierarchical scenario, the LF is predicted to have a steep faint-end slope, reflecting the increasing number of low mass systems with faint magnitudes \cite{wh91}. Therefore, studies of the LF, both in the field and in clusters, is one of the most direct probes for testing the hierarchical model.

Detailed studies of the spatial and morphological distribution of cluster galaxies are important for understanding the internal properties of clusters, as well as the evolution of these over time. To do this observationally, it is necessary to isolate the galaxies in a given field belonging to the cluster, from the foreground or background field galaxies, projected on to the cluster. Spectroscopy is costly, since a cluster field may consist of thousands of galaxies. More important, for clusters at higher redshifts spectroscopy is limited to only the most luminous galaxies. The traditional way of correcting for field contamination at faint magnitudes is to use a cluster-free field, preferentially close to the cluster, and calculate the number of galaxies as a function of magnitude to be subtracted from the cluster image. This method is, however, increasingly uncertain at higher redshifts, where the ratio of field galaxies to cluster galaxies increases (e.g. Driver at al. 1998a). In Section 3.7 we show that the field contamination can exceed 80 per cent of the total number of galaxies for clusters at redshifts $z~\gsim~0.5$. A further important disadvantage is that it is not possible to assign membership to a specific galaxy with any confidence.

An alternative and more powerful way to determine cluster membership is to use photometric redshifts based on broad band photometry to estimate redshifts. The main advantage with this method is that a large number of redshifts can be determined in a relatively short time, and to magnitudes considerably fainter than traditional spectroscopy. An important bonus is that the photometric redshift technique also provides an estimate of the Hubble type for the observed galaxies, or more precisely, their broad spectroscopic characteristics. This information can be used, together with redshift information, to determine the fraction, as well as the spatial distribution, of different types of galaxies. We later discuss this technique in more detail.

The trade-off with this method is that uncertainties in the estimated redshifts are considerably larger compared to spectroscopy. The accuracy has, however, improved and typical rms deviations between spectroscopic redshifts and broad band photometric redshifts are $\Delta z_{rms}\sim$ 0.05 at $0.6~<~z~<~1.0$ \cite{co95,li98,lu99} and $\Delta z_{rms}/(1+z)=$ 0.06--0.1 up to $z~\sim~6$ \cite{fe99,be00}. This accuracy is sufficient for determination of cluster membership and absolute magnitudes, but is clearly not useful for dynamic studies.

In this paper we demonstrate the use of photometric redshifts for a population study of the cluster Cl1601+42. With information on cluster membership and spectroscopic galaxy type, we calculate the luminosity function separately for early-type and late-type galaxies within the cluster, as well as their spatial distribution. Previously, photometric redshifts have been used to determine cluster membership for a few clusters only. The most extensive study of photometric redshifts applied to clusters of galaxies is presented by Brunner \& Lubin \shortcite{br00}. They in paricular, discuss how the method can be used to either maximize completeness, or to minimize contamination from field galaxies. Lubin \& Brunner \shortcite{lu99} present results on the distribution of galaxy morphologies in two distant clusters, where cluster membership is determined by photometric redshifts. Kodama, Bell \& Bower \shortcite{ko99} use photometric redshifts for determining cluster members in Abell 370 at $z~=~0.37$. Connolly et al. \shortcite{co96} include photometric redshifts when investigating a supercluster structure at $z$ = 0.54, and Connolly, Szalay \& Brunner \shortcite{co98} use the method for measuring the evolution of the angular correlation function in the HDF. Finally, B\'{e}zecourt et al. \shortcite{bz00} use photometric redshifts when analysing weak lensing of the cluster Abell 2219 at $z$ = 0.22. For studies of the LF and population content of clusters this method has, however, only had limited use.

The cluster Cl1601+42 is an intermediate rich cluster at $z$ = 0.54. It was chosen because there already exist both high resolution HST imaging of the core of the cluster \cite{sm97}, and a fairly large number of spectroscopic redshifts of galaxies in the cluster field \cite{dr99}. The cluster is included in the MORPHS sample of ten clusters at $0.37~<~z~<~0.56$ (Smail et al. 1997). The available information is valuable both for testing the photometric redshifts technique, as well as for determining the morphological properties of the cluster.

The plan of this paper is as follows. In Section 2 we present the data, in Section 3 we describe the photometric redshift technique used. Results are presented in Section 4, followed by a discussion in Section 5. Finally, conclusions are given in Section 6.
 
We assume in this paper a flat cosmology with $\Omega_M=0.3$ and $\Omega_{\Lambda}=0.7$, and a Hubble constant $H_0=50$ km s$^{-1}$ Mpc$^{-1}$. The absolute magnitude for other choices of the Hubble constant are related by: $M_{(H_0=50)} = M_{(H_0=65)} - 0.57 = M_{(H_0=100)} - 1.51$. At the cluster redshift, $z$ = 0.54, the absolute magnitudes for different cosmologies are related by: $M_{(\Omega_M=0.3, \Omega_{\Lambda}=0.7)} = M_{(\Omega_M=1.0, \Omega_{\Lambda}=0.0)} - 0.42 = M_{(\Omega_M=0.2, \Omega_{\Lambda}=0.0)} - 0.24$.

Magnitudes are given in the standard Vega based system.
\section{The Data}
\subsection{The cluster Cl1601+42}
Cl1601+42 is located at R.A. 16$^h$03$^m$09$^s$.8 and Dec. $42^{\circ}$~45\arcmin~18\arcsec (J2000.0). The redshift has been determined spectroscopically to $z~=~0.54$ \cite{ok96}, and the velocity dispersion is 1210 km s$^{-1}$ \cite{dr99}. The cluster is a weak X--ray source with $L_x$ (0.3--3.5 keV) = 0.35 $h^{-2}$ 10$^{44}$ ergs \cite{he82,sm97}. The galactic extinction is consistent with $A_B~=~0.00$ \cite{sc98}. 
\subsection{Observations}
\begin{figure*}
\begin{minipage}{140mm}
\psfig{figure=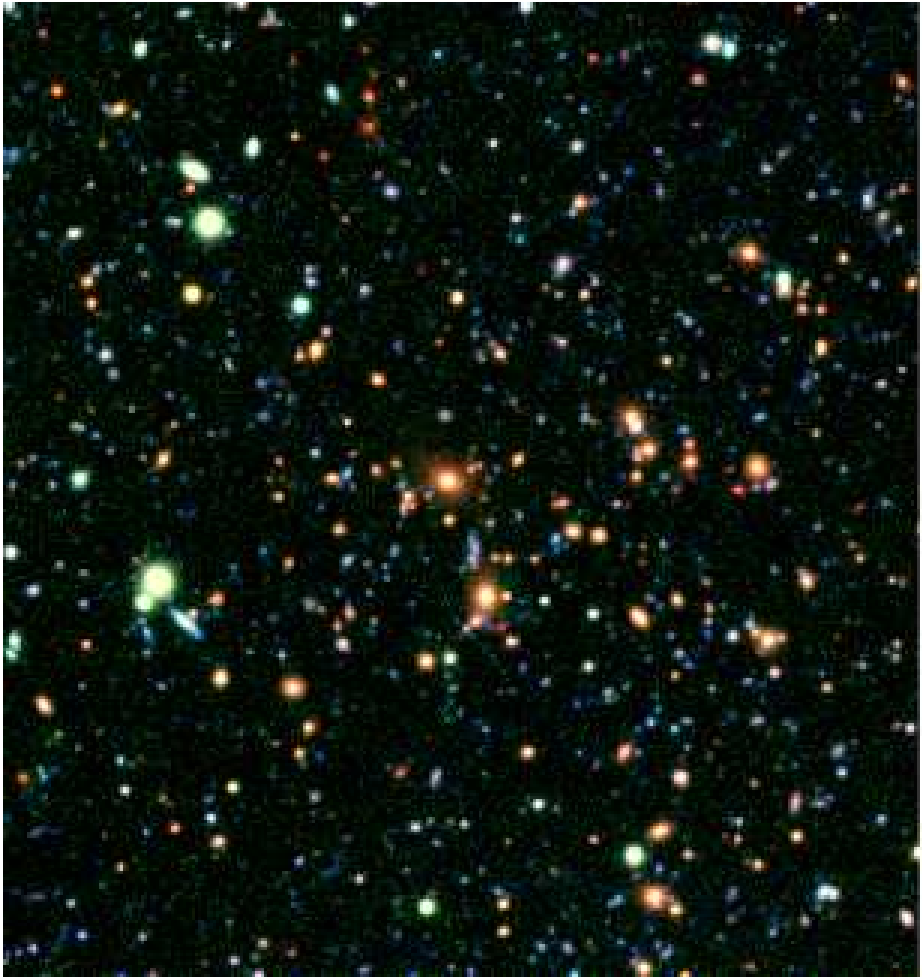,height=13cm}
\caption{Image of Cl1601+42 covering an area 3\arcmin.7 $\times$ 3\arcmin.7, which is $\sim$ 50\% of the total field covered.}
\label{Figure1}
\end{minipage}
\end{figure*}

The cluster was observed with the 2.56m Nordic Optical Telescope (NOT) and the Andalucia Faint Object Spectrograph and Camera (ALFOSC) during four observational runs 1997--1999 (Table \ref{Table1}). ALFOSC contains a Loral 2K$^2$ chip with a field of 6\arcmin.5$\times$6\arcmin.5 and an image scale of 0$\arcsec$.188/pixel. For the observations during 1997 we used the NOT CCD\#6 and for the 1998--99 observations the NOT CCD\#7. Cl1601+42 was observed in five filters, $U$, $B$, $V$, $R$ and $I$. In the $R$ band we also observed a neighbouring blank field, which we refer to as the background field. The position of this field is R.A.~16$^h$08$^m$54$^s$.0 and Dec.~$41^{\circ}$~34\arcmin~00\arcsec (J2000.0). All observations were performed under photometric conditions. The seeing in the images varies between 0\arcsec.75 and 1\arcsec.0, depending on filter. A complete log of the observations is given in Table \ref{Table1}. Fig. \ref{Figure1} shows a colour image of Cl1601+42, constructed from the $B$, $R$ and $I$ bands. 
\subsection{Data reduction}
The data were reduced using the {\sc iraf} and {\sc midas} packages. Bias subtraction was made in a standard manner. Because of the large number of science frames in each filter, flat-fields were constructed directly from these images after removing objects. Because the $I$--band science frames show a fringe pattern we used short exposure, twilight frames for the flat-field. After flat-fielding the science $I$--images, a fringe frame was created, which was subtracted from the science images. The flat-fielded images in each filter were corrected for atmospheric extinction, aligned, and finally combined. The final size of the field is 5\arcmin.5 $\times$ 5\arcmin.6 arcmin. The images were calibrated using standard stars from Landolt \shortcite{la92} and Christian et al. \shortcite{ch85}. We estimate the photometric errors in the zero point calibration to $\sigma_m~\lsim~0.05$ in $U$, $B$, $V$ and $R$, and $\sigma_m~\sim~0.07$ in $I$.
\begin{table*}
\begin{minipage}{140mm}
\caption{Log of observations.}
\begin{tabular}{lclrcc}
\bf Object &\bf Filter & \bf Obs. date & \bf Exp. time & \# of exp. & Seeing \\ 
& & & & & \\
Cl1601+42 & $U$ & Apr/Jun 99 & 25200s & 14 & 0\arcsec.89 \\ 
 & $B$ & Jun 98 & 12600s & 14 & 0\arcsec.84 \\ 
 & $V$ & Apr 99 & 16200s & 18 & 1\arcsec.02 \\ 
 & $R$ & Jun 97 & 7200s & 8 & 0\arcsec.75 \\
 & $I$ & Jun 98/Jun 99 & 14400s & 24 & 0\arcsec.76 \\ 
Background & $R$ & Jun 97 & 7200s & 8 & 0\arcsec.75 \\
\end{tabular}
\label{Table1}
\end{minipage}
\end{table*}
\subsection{Object catalogue}
The reduced images were analysed using the {\sc focas} package \cite{ja81,va82,va93}. We used a value of 3$\sigma$ of the sky noise as a limit to our detections, and a minimum detection area corresponding to the seeing, ranging from 18 pixels in $R$ to 30 pixels in $V$. The ordinary {\sc focas} procedure, including sky correction and splitting of multiple objects, were used. For each filter we made a catalogue including isophotal magnitude, position and area. In order to calculate colours of the objects we also measured the magnitude using a constant aperture for each object. Before doing this, we smoothed each image to a seeing corresponding to that of the $V$ image, which had the worst seeing. The diameter of the aperture corresponds to twice the FWHM of the smoothed images, i.e ten pixels. A combined catalogue was made from the positions in the $R$--catalogue and matching these to the positions in the other catalogues. A match was considered satisfactory if the distance between the centre positions of two objects in the different catalogues was less than 9 pixels. This distance was chosen to be large enough that each specific galaxy is found in as many colours as possible, but sufficiently small so that no misidentifications occur. To check this, we visually inspected cases near the limit. The average offset between galaxies in the different colours is $\lsim$ 2 pixels.

\subsection{Photometric limits, completeness, magnitude corrections and errors}
The 1$\sigma$ limiting surface brightness for the different filters are, in $mag$ $arcsec^{-2}$, $\mu_U$ = 26.7, $\mu_B$ = 27.5, $\mu_V$ = 27.3, $\mu_R$ = 26.8 and $\mu_I$ = 26.2.  

To estimate the magnitude correction that should be applied when calculating the total magnitudes from the observed isophotal magnitudes we used the same method as in N\"{a}slund, Fransson \& Huldtgren \shortcite{na00} (see also Trentham 1997). This correction accounts for the loss of light from the part of the galaxy profile that is outside the limiting isophote. For galaxies with exponential profiles, the correction is $0.04~\lsim~\Delta m~\lsim~0.10$ at $R~=~25$, whereas the correction is negligible at $R~<~24$. For galaxies with a de Vaucouleurs profile, the correction is $\Delta m~\sim~0.1$ for galaxies with $R~<~24$. We assume that only galaxies determined as early-types (a definition is given in Section 3.5) with $R~\lsim~24$ have de Vaucouleurs profiles, while fainter early-types are assumed to have exponential profiles (see discussion in Trentham 1997). We hereafter include these corrections when quoting magnitudes.

To estimate the completeness, we added artificial galaxies with different magnitudes, radial profiles, scale lengths and inclinations to the images, and used {\sc focas} with the same settings as for the real data to detect these objects. At $R~=~25$, which we in the following use as our completeness limit for the photometric redshift sample, we recover all of the simulated galaxies. At $R~=~25.5$ we recover $\sim~97$ per cent, and at $R~=~26$ between 69--85 per cent of the simulated galaxies, depending on the galaxy type.

For bright objects the magnitude errors are dominated by the zero point calibration errors (given above). For faint objects the errors increase due to lower signal-to-noise ratio. At the limit, $R~\sim~25$, we estimate the total error in $R$ to be $\sigma_m~\sim~0.10$.
\section{Photometric redshifts}
\subsection{Method}
There are basically two different approaches to obtain photometric redshifts. The empirical fitting method (e.g. Connolly et al. 1995; Connolly et al. 1997; Wang, Bahcall \& Turner 1998; Brunner, Connolly \& Szalay 1999) uses colours and magnitudes of objects with known spectroscopic redshifts to calculate a relation in form of a polynomial between these quantities. The observed magnitudes from galaxies with unknown redshifts are then inserted in this polynomial in order to calculate the photometric redshift. The alternative template fitting method (e.g. Puschell, Owen \& Laing 1982; Gwyn 1995; Mobasher et al. 1996; Sawicki, Lin \& Yee 1997; Fern\'{a}ndez--Soto et al. 1999; Bolzonella, Miralles \& Pell\'{o} 2000; Fontana et al. 2000) uses observational or theoretical galaxy templates of the whole range of Hubble types, shifted to a given redshift. A $\chi^2$--fit is made that minimizes the difference between the observed and the predicted fluxes, magnitudes or colours. 

The advantages and disadvantages of the different methods are discussed in e.g. Hogg et al. \shortcite{ho98}, Koo \shortcite{koo99} and Bolzonella et al. \shortcite{bo00}. A major drawback with the empirical fitting method is that it requires a relatively large set of spectra over the whole redshift range being probed. This is difficult, or impossible, for the faintest galaxies, and can easily introduce systematic errors for these. 

The weak point of the $\chi^2$--method is the template set used. Redshifted templates from locally observed galaxies are most often used to represent galaxies at earlier epochs. If there are significant evolutionary changes in spectral shape between galaxies today and at higher redshifts, this may introduce systematic errors. A lower metallicity may have the same effect. Dust extinction is another source of uncertainty. 

Synthetic spectra are affected by similar effects. Internal dust absorption can be included by having $E(B-V)$ and the extinction law as free parameters when constructing the template set. Effects of different metallicities can also be taken into account (e.g., Furusawa et al. 2000; Bolzonella et al. 2000), as well as absorption due to intergalactic \hbox{H\,{\sc i}} clouds \cite{ma95}. The latter is, however, mainly a concern for high-$z$ searches, since the Lyman break is redshifted into the $U$--band only at $z~\gsim~2$.

Here we employ a version of the template fitting method. This is the most commonly used method and, despite the possible risk for systematic errors caused by an insufficient knowledge of high-$z$ galaxy templates, it has been shown to be very successful. In the redshift range we are interested in, i.e., 0 $<z~\lsim~1.5$, the errors are typically $\Delta z_{rms}/(1+z)~\approx ~0.05$ \cite{co00}, depending on the number of filters used and the accuracy of the photometry. The spectroscopic redshifts available for the area we have observed serve as a check on the accuracy of the derived photometric redshifts. 

For every object we minimize the expression
\begin{equation}
\chi^2(t,z,m_{\alpha})=\sum_{i=U,B,V,R,I}\frac{(m_i-(T_i(t,z)+m_{\alpha}))^2}{\sigma_i^2}
\end{equation}
where $m_i$ and $\sigma_i$ are the observed magnitudes and errors in the different filters, respectively. $T_i(t,z)$ is the magnitude in the $i$--filter of template $t$, redshifted to $z$. This magnitude is calculated by convolving the filter curve and the quantum efficiency of the detector with the galaxy template. The quantity $m_{\alpha}$ is a constant, which fits the apparent magnitude of the template galaxy. 

A set of 20 different templates are used. We construct ten of these by interpolations between the four observed galaxy templates given by Coleman, Wu \& Weedman \shortcite{co80}. These represent E, Sbc, Scd and Im galaxies. Further, Sa and Sb templates from the Kinney--Calzetti spectral atlas of galaxies \cite{ki96} are added to these. Finally, we have constructed eight templates representing E+A galaxies, using the Galaxy Isochrone Synthesis Spectral Evolution Library (GISSEL) \cite{br93,br95}. These templates are characterised by an old population ($\sim$ 8 Gyr) to which a starburst of duration 0.25 Gyr is added. The mass of the starburst is set to be 20 per cent of the total mass. The eight templates represent galaxies at epochs 0.5 Gyr to 3 Gyr after the burst. The parameters used to construct the E+A galaxies are chosen to be close to the preferred model used by Belloni et al. \shortcite{be95}, who compare colours from the templates with colours from a combination of broad and narrow band observations of E+A galaxies.

The inclusion of E+A galaxies are more important for clusters of galaxies than the field. Belloni et al. \shortcite{be95} and Dressler et al. \shortcite{dr99} show that $\sim$20 per cent of the brighter cluster galaxies at $0.4~\lsim~z~\lsim~0.6$ have E+A features, while only a few per cent of the field galaxies show these characteristics. Absorption due to intergalactic \hbox{H\,{\sc i}} clouds is treated as in Madau \shortcite{ma95}. With these templates our sample include a fair distribution of galaxy types, including starburst and post-starburst galaxies. Each galaxy template is redshifted in steps $\Delta z~=~0.002$ from $z~=~0$ to $z~=~2.0$. Besides the galaxies, we also include eight stellar templates of M--dwarfs taken from Gunn \& Stryker \shortcite{gu83}.  
\subsection{Accuracy of the photometric redshifts}
\begin{figure}
\psfig{figure=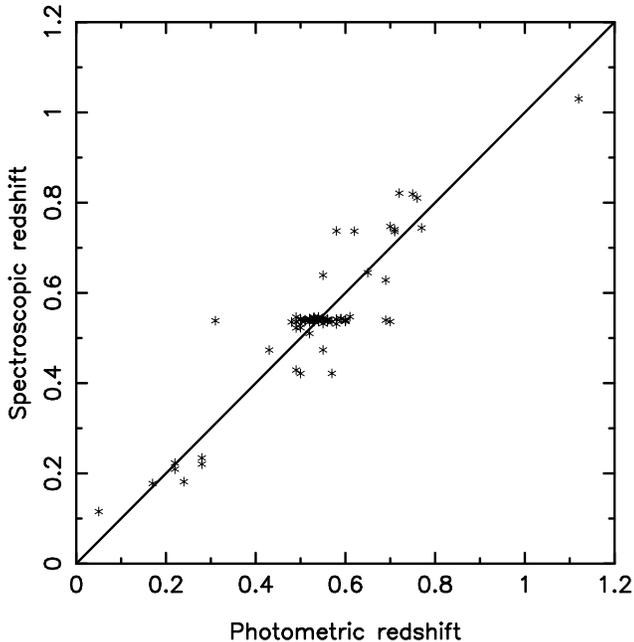,height=8.5cm,angle=-90}
\caption{Photometric versus spectroscopic redshifts for 78 galaxies in the field containing Cl1601+42. The cluster is located at $z=0.54$. The deviation between the spectroscopic and the photometric redshifts is $\Delta z_{rms}\sim$0.08. The straight line shows the location of $z_{phot}=z_{spec}$. Spectroscopic redshifts are taken from Dressler et al. (1999).}
\label{Figure2}
\end{figure}

Dressler et al. \shortcite{dr99} have obtained 102 spectroscopic redshifts in the field of Cl1601+42. Of these, 58 are considered to be cluster members, 40 belong to the field and four are stars. To check the accuracy of our photometric redshifts, we have compared our determinations with this spectroscopic sample. We have excluded stars, galaxies lying outside our field, galaxies with spectroscopic redshifts that are not considered as certain (as commented in the catalogues by Dressler et al.) and two galaxies that appear twice in the MORPHS catalogues. Finally, we have excluded two galaxies where the photometric redshifts together with inspection of the spectra (published by Dressler et al.) indicates that the redshifts given in the MORPHS catalogues are incorrect. Fig. \ref{Figure2} shows photometric versus spectroscopic redshifts for the remaining 78 comparison galaxies. The straight line in the figure shows the location of $z_{phot}=z_{spec}$. The rms deviation in redshift is $\sigma_z\sim$ 0.076, and a fit to the data points gives $z_{phot}~=~(0.98\pm~0.02)\times~z_{spec}$.

We calculate photometric redshifts for all object that are detected in at least three filters. If an object is not detected in all filters, we use this non-detection to set an upper limit to the magnitude, which is used in the redshift determination. Of the 79 galaxies with spectroscopic redshifts we have magnitudes in all five filters for 63 objects, whereas 13 objects are detected in four filters and three objects in three filters only. There is no difference in accuracy between the subsamples with detection in four and five filters. In all cases the $U$--band magnitude is lacking.

Liu \& Green \shortcite{li98} discuss the importance of the $U$--band for photometric redshift determinations. In their sample they find that $\Delta z_{rms}$ doubles if they exclude the information from the $U$--band, compared to when they include all their six filters. They note that one of the main causes for this larger scatter is misclassification of Scd and Irr galaxies at $z~\sim~0.3$ with Sbc galaxies near $z~\sim~0.05$. At redshifts $z~\gsim~0.4$, where the 4000~\AA~break moves into the $V$--band, the importance of $U$--band information is decreased. To confirm this we have tested this case by excluding the $U$ magnitude for the 63 objects with detections in all filters. The r.m.s. deviation is now $\sigma_z~\sim~0.11$, compared to $\sigma_z~\sim~0.076$ with five filters. For spectroscopically determined cluster members the accuracy is $\sigma_z~\sim~0.06$ with five filters, and $\sigma_z~\sim~0.08$ with four filters.

Because we are mainly interested in accurate redshifts near the cluster redshift, i.e. at $0.4~\lsim~z~\lsim~0.7$, the possible misclassification above should not influence the cluster population, and we therefore include also galaxies with only $BVRI$ photometry.

Galaxies with detection in three filters have $\sigma_z\sim$ 0.05. The sample does, however, only consist of three galaxies and is therefore probably not representative. To check the three filter case, we instead determined redshifts for the whole spectroscopic sample where we only use the information in $BRI$. If we exclude the $V$ filter, which after the $U$ filter, is the filter reaching the smallest depth, we obtain $\sigma_z\sim$ 0.16. The increase is mainly caused by a few outliers. For spectroscopically determined cluster members $\sigma_z\sim$ 0.13 with three filters. Excluding three outliers results in $\sigma_z\sim$ 0.09.

Without $IR$ photometry the photometric redshifts in the interval $1~\lsim~z~\lsim~2$ should be less accurate, since here the 4000~\AA~break has moved long-ward of the $I$--band, and there are no other distinct features to detect. Including JHK photometry allows for the detection of the 4000~\AA~break in this interval. Fontana et al. \shortcite{fo00} and Furusawa et al. \shortcite{fu00} compares results obtained by optical four--band and optical+$NIR$ 7--band photometry of galaxies in the HDF and find that the accuracy is considerably worse in the interval $1~<~z~<~2$ when using optical filters only. Below $z~\sim~1$ the influence of the $NIR$ bands is, however, small. Since we are primarily interested in redshifts at $z~\sim~0.5$, our results should not be much affected by the lack of $NIR$ photometry.

\subsection{Systematic effects}
Besides the statistical uncertainty affecting the redshift from which
the cluster membership is determined, there may be systematic errors
in the redshift determination that may depend on increasing
photometric errors, or even non-detections in some bands, at faint
magnitudes, and between different galaxy types. 

As shown in Section 3.2, the dispersion between spectroscopic and photometric redshifts increase when photometry is lacking in one or two filters. To examine the magnitude dependence of this effect, we calculate for each magnitude bin the mean dispersion, where we take into account the larger errors for galaxies that are not detected in all filters. In the faintest magnitude bin
($24.5~<~m_R~<~25$) we find that the dispersion increases by $\Delta
\sigma_z~<~0.006$ compared to the over all mean. For a Gaussian
distribution of the dispersion, the extra fraction of galaxies lost in the
faintest bin is $<~3$ per cent, clearly less than statistical
errors. The reason for this is that we have chosen a magnitude limit
that is sufficiently bright for galaxies with photometry in all five
filters to dominate even in the faintest bin.

Systematic effects can also be introduced because errors in the redshift
increase at faint magnitudes due to the larger photometric
errors. To quantify this we have made Monte Carlo simulations, where
we create a catalog with $10^4$ cluster galaxies. Each galaxy is placed
at the cluster redshift and is given magnitudes in the different bands
from the spectral templates. To each band we add photometric
errors corresponding to the magnitude dependence of the real
data. We now determine the photometric redshift for each
galaxy, and compare with the input galaxy redshift. 
%We find a small,
%but significant, increase in the errors of the photometric redshifts
%at fainter magnitudes due to the larger photometric
%uncertainties. 
In the faintest bins ($m_R~\gsim~24$) we find that $\sim~2$ per cent
are outliers, i.e. have errors that are $\Delta z~\gsim~0.5$. For
Cl1601+42 the number of galaxies lost due to this effect is at most a
few. Secondly, and more important, we find that the increase in errors
for the remaining galaxies in the faintest bins is $\Delta
\sigma_z~\sim~0.01$, which translates to a loss of $\sim~5$ per cent.

Using the same MC simulations we can study the error dependence on
galaxy type. Due to the flatter spectrum of the late-type galaxies,
the error in the redshift is slightly larger compared to early-types,
which have a more distinct 4000 \AA~ break. The fraction of
galaxies lost should therefore be higher for late-types compared to early
types. All outliers in the faintest magnitude bins are consequently
found to be of late-types.

The magnitude limit of the Dressler et al. spectroscopic sample, which
is used to calculate the errors in the photometric redshifts, is
$R~\sim~23.3$. If high-$z$ galaxies fainter than this have spectral
energy distributions that are not represented by the range of
templates used, there could be systematic effects affecting the
accuracy. The comparison of the spectroscopic and photometric
redshifts in HDF North, published by Fern\'{a}ndes--Soto et
al. \shortcite{fe99}, does, however, not show any decrease in accuracy
for the photometric redshifts in the magnitude range
$23.5~\lsim~R~\lsim~25$, compared to the range $R~\lsim~23.5$ at
redshifts $0.5~<~z~<~2.0$. This argues that there should be no
dramatic change of the spectral energy distribution of the faint
galaxies, which is not covered by our templates, unless faint {\it
cluster galaxies} have a spectral shape different from faint field galaxies.

Summarizing, we find that there is a slight systematic magnitude
dependence of the errors in the photometric redshifts, and that errors
depend on galaxy type. However, the systematic effects introduced by
these effects are unlikely to dominate over the statistical
errors. The systematics present should lead to a modest underestimate of
the number of faint galaxies, which should mostly affect late types.

\subsection{The photometric redshift catalogue}
For the objects with photometry in at least three filters we have constructed a photometric redshift catalogue. After excluding 83 objects classified as stars, the catalogue consists of 1568 objects.

The catalogue contains photometric redshifts for 1140 galaxies to $R~=~25$, which is 95 per cent of all galaxies to this limit. Of these, 807 redshifts are based on photometry in five filters, while 227 and 106 redshifts are based on four and three filters, respectively. 
\subsection{Determination of galaxy types}
A classification into morphological types based on visual (or automatic) inspection requires high resolution imaging, such as obtained from the HST. This is for our cluster only available for the central $\sim$ 5 square arcmin. Even without high resolution imaging, colours can be used to estimate the galaxy types. This is basically what the template fitting method does when it compares observed magnitudes with magnitudes derived from a set of spectral galaxy templates. There is, however, not a one-to-one correspondence between the galaxy type derived from colours and the visually determined morphological types. Examples are red spirals with little ongoing star formation \cite{po99}. Further, van Dokkum, Franx \& Kelson \shortcite{va98} find that early-type galaxies, especially faint S0 galaxies, can have relatively blue colours. Also, galaxies with E+A type spectra can not be classified into any specific morphological type. Poggianti et al. \shortcite{po99} note that galaxies in this class mainly have disc-like morphologies, but examples of all Hubble types are present. For the future discussion it is important to keep in mind that galaxy classification by the photometric redshift method is basically a spectroscopical classification, rather than a morphological. 

In this study we divide galaxies into two broad categories, based on the spectral class determined by the photometric colours of the galaxies. The division is made half way between the E and Sbc templates from Coleman et al. (1980), approximately corresponding to a rest frame colour $B-V~=~0.8$. The first category consists of early-type galaxies with red colours, which are mostly ellipticals and lenticulars, but also include passive spirals with red colours. The second category includes late-type spirals and irregulars, with a possible inclusion of blue elliptical like systems. This division is the same as used by Ellingson et al. \shortcite{el01}, when separating red and blue galaxies.

\subsection{The redshift distribution}
\begin{figure*}
\begin{minipage}{140mm}
\psfig{figure=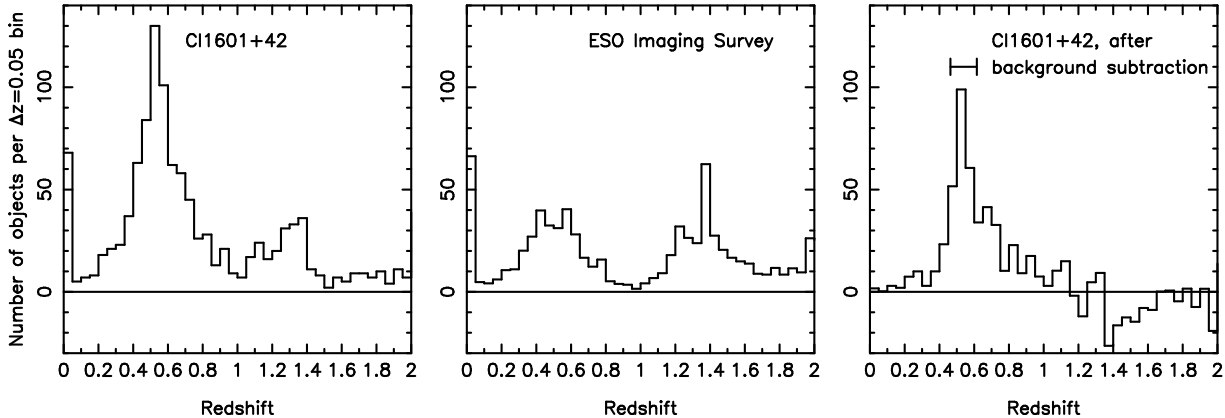,height=5.5cm,angle=-90}
\caption{Left panel: Distribution of photometric redshifts for 1140 objects in the field containing Cl1601+42 to magnitude $R$ $<$ 25. Middle panel: Same for the ESO Imaging Survey, corrected to the same field size as the Cl1601+42 field. Right panel: Residuals after subtracting the background field from the Cl1601+42 field. The redshift range chosen for the cluster analysis is marked by the horizontal bar.}
\label{Figure3}
\end{minipage}
\end{figure*}

%\subsection{Redshift distribution of galaxies in the cluster field}
The redshift distribution of the objects in the field of Cl1601+42 is shown in the left panel of Fig. \ref{Figure3}. In total, 1140 galaxies with $R$ $<$ 25 are included. There is a clear peak around the cluster location at $z\sim$ 0.54, superimposed on a broader distribution of field galaxies with redshifts in the range $0.3~\lsim~z~\lsim~0.7$. A small secondary maximum in the distribution of field galaxies is apparent at $z\sim$ 1.3. This is likely to be caused by our lack of $NIR$ bands, since we above $z\sim$ 1.2 are no longer able to detect the 4000~\AA~break.

A relatively large fraction of the objects in the lowest redshift bin is probably stars. We have already excluded 82 objects from this bin that are best fitted by M--star spectra, or have a PSF of a point source. The magnitudes of the remaining objects in the bin are, however, too faint to allow a secure classification from an inspection of the PSF. The lowest bin may therefore contain a number of stars in addition to low redshift galaxies.

To estimate the number of field galaxies to our limiting magnitude, we have calculated photometric redshifts for the objects in the ESO Imaging Survey (EIS) \cite{da98}. Using aperture photometry in $U$, $B$, $V$, $R$ and $I$ we have determined photometric redshifts for 707 objects to $R~=~25$ in the 5\arcmin.3 $\times$ 5\arcmin.3 EIS-field including the HDF--South area. The redshift distribution, scaled to the area of Cl1601+42, of these objects is shown in the middle panel of Fig. \ref{Figure3}. Besides the peak in the distribution at $0.4~\lsim~z~\lsim~0.7$, there is a second peak at $z~\sim~1.4$, similar to the Cl1601+42 field. Since for consistency we only include optical filters for the EIS data, this peak may also be caused by the lack of $NIR$ photometry.

Field-to-field variations caused by large scale structures introduce an uncertainty when correcting for field galaxies. To check if the EIS field has a number density that is comparable to other fields, we use estimates from Smail et al. \shortcite{sm97}, who find that the Medium Deep Survey \cite{gr94} has a field galaxy density of $\sim~11.2$ galaxies per arcmin$^2$ to $I~=~22.6$. From the EIS field we find $\sim~11.0$ galaxies per arcmin$^2$ to the same magnitude limit suggesting that the background counts in the EIS field are normal. Deviations may, however, increase at fainter magnitudes. To quantify the field-to-field variations, we calculate the variance in the background counts from the angular correlation function (Peebles 1975; 1980). At $R~>~24$ we use a correlation amplitude from Connolly et al \shortcite{co98}, while for brighter magnitudes we calculate the correlation amplitude from a fit to the data in Woods \& Fahlman \shortcite{wo97}. The 1$\sigma$ variation is found to be $\pm~$ 13 per cent at $R~\sim~25$ for our field size.
\subsection{Cluster membership}
The limited accuracy of the photometric redshifts makes the determination of cluster membership somewhat complicated. First, a redshift range defining the cluster has to be determined. Then, the contamination of field galaxies within this redshift range should be estimated. 

For the spectroscopic sample, Dressler et al. \shortcite{dr99} use a redshift range $0.5100~\le~z~\le~0.5474$ to define cluster membership. In the photometric sample the higher dispersion in redshift has to be taken into account, and a broader interval must be used. The redshift interval has to be large enough so that as many of the cluster galaxies as possible are retained, but at the same time keeping the number of field galaxies in the cluster redshift range to a minimum. Here, we take a range around the mean redshift of the cluster corresponding to 1.5$\sigma_z$, 
%i.e $z_{Cl}=z_{mean}\pm ~1.5\sigma_z$, 
corresponding to $z_{Cl}$= 0.43--0.65.
The interval is shown in the right panel of Fig. \ref{Figure3}, and contains 418 galaxies. Correcting for the few percent of galaxies with $R~<~25$ that lack photometric redshifts results in a total of 432 galaxies. After renormalising for the different size of the EIS field, we find that 185 field galaxies are expected within the above redshift range. The total number of cluster galaxies after subtracting the background within the cluster redshift range is therefore 247.

The 1.5$\sigma$ cut in the redshift range means that we deliberately loose some cluster galaxies. Increasing the redshift range would, however, increase the background contamination, and therefore increase the uncertainty. In particular, the probability that a specific galaxy is a cluster member would decrease. For a Gaussian distribution of the redshift errors we estimate that $\sim~20$ per cent of the cluster population is outside the redshift range. Here, we have assumed that the r.m.s. error in the catalogue is higher than in the comparison with the spectroscopic sample by $\Delta\sigma~=~0.01$, since a slightly larger fraction of galaxies in the catalogue do not have photometry in all five filters, as compared to the spectroscopic sample. When appropriate, we correct for this loss of galaxies in the analysis.

The difference between using photometric redshifts and an ordinary background subtraction is that we only have to do a subtraction for a limited redshift range. This decreases the errors compared to using the full range of redshifts. A simple estimate, based on Poissonian statistics, illustrates this. Of the total number of $\sim~1200$ galaxies we estimate that 247 are cluster galaxies and the remaining $\sim~950$ are field galaxies (for simplicity we do here not correct for galaxies lost due to the 1.5$\sigma$ cut). With a standard subtraction method, the r.m.s. error in the number of cluster members is equal to the r.m.s. error in the subtracted background, $\pm \sqrt{950}$. This makes the number of cluster galaxies $247~\pm~31$. When only subtracting the $\sim$ 185 field galaxies within the cluster redshift range the r.m.s. error is reduced to $\pm~\sqrt{185}$, and the number of cluster galaxies becomes $247~\pm~14$. The r.m.s. error is therefore reduced by more than a factor two. 

Most important is that the probability that a specific galaxy is a cluster member increases from 247/1200 $\sim$ 21 per cent without redshift information to 247/432 $\sim~58$ per cent with redshift information. With a field contamination as high as $\sim~80$ per cent, it is impossible to obtain useful information about e.g. the radial distribution within the cluster in the standard subtraction method. With the limited subtraction necessary when using the photometric method, we can, however, derive meaningful information about the internal properties of the cluster. In combination with the rich colour information of the individual galaxies, this method allows more complex investigations of the cluster properties. A further advantage is that the field-to-field variations are minimized.

\section{Results}
\subsection{Cluster properties}
Abell \shortcite{ab58} defines {\it Richness classes} by the number of galaxies in the cluster with magnitudes between $m_3$ and ($m_3$ + 2), where $m_3$ is the third brightest galaxy in the cluster. Here we find $m_3~\simeq~20.1$ in the $R$--band, corresponding to $M_B~\simeq~-22.5$. We find 135 galaxies in this interval. After subtracting for background galaxies within the redshift range of the cluster (see below), we estimate the resulting number of cluster galaxies to $\sim~70$. This makes Cl1601+42 a cluster of Abell class 1--2, i.e. an intermediate rich cluster. This is important to keep in mind when comparing with other results, since most investigations concern rich clusters. 

Clusters can also be characterised by the degree of central concentration. Butcher \& Oemler \shortcite{bu78b} define a concentration index, $C$ = log$(R_{60}/R_{20})$, where $R_n$ is the radius of a circle containing $n$ per cent of the projected galaxies in the cluster with $M_V~<~-20$ (using a cosmology with $q_0~=~0.1$ and $H_0~=~50$ km s$^{-1}$ Mpc$^{-1}$). BO84 define clusters with $C~\approx~0.3$ as open or irregular. Clusters with $C~\gsim~0.4$ are normally referred to as compact or concentrated. 

We estimate that Cl1601+42 has $C~\sim~0.40\pm~0.04$. Because our field does not contain the whole cluster, we have extrapolated the radial profile according to an isothermal sphere outside of our image. This makes the value of $C$ somewhat uncertain. Dressler et al. \shortcite{dr97} find $C~=~0.34$ for Cl1601+42. The difference in $C$, if significant, is probably due to different methods used for calculating the background contamination. Of the clusters in the MORPHS sample, Cl1601+42 belongs to the group with relatively low concentration. 

\subsection{The cluster luminosity function}
%\subsubsection{The luminosity function}
\subsubsection{The photometric redshift method}
\begin{figure}
\psfig{figure=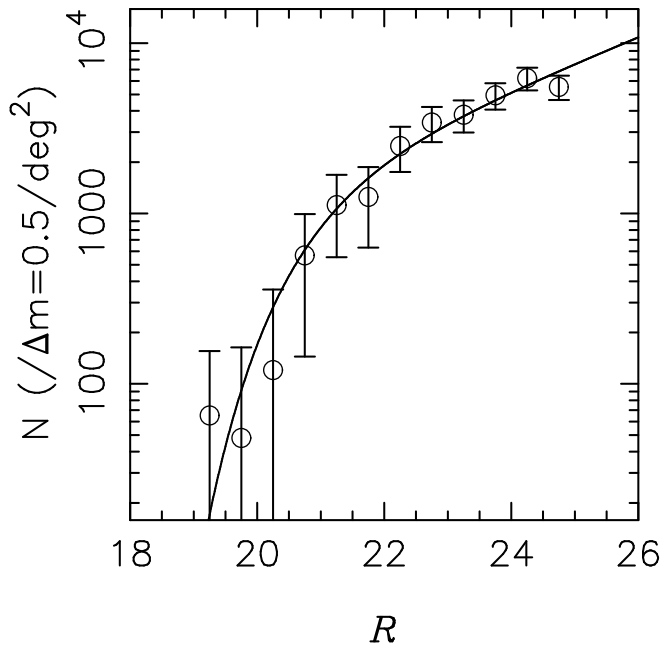,height=8.5cm,angle=-90}
\caption{Luminosity function in the $R$--band for Cl1601+42. Error-bars include Poissonian uncertainties and field-to-field variance.}
\label{Figure4}
\end{figure}

To calculate the LF we divide the 418 galaxies within the cluster redshift range into magnitude bins of $\Delta m~=~0.5$ to $R~=~25$. We smooth each galaxy count with a Gaussian using a $\sigma$ corresponding to the error in magnitude. The effect of this is small since the errors are typically much less than the bin size. It does, however, result in bins having fractions of galaxies, which can be noted in the brightest bins. In the faintest magnitude bins we correct for the fraction of the galaxies that lack photometric redshifts by multiplying the number of galaxies with the inverse of the completeness of the photometric catalogue. After correcting for field galaxies in the redshift range of the cluster for each magnitude bin, we obtain the LF shown in Fig. \ref{Figure4}. The error--bars in the figure represent 1$\sigma$, including Poissonian uncertainties and field-to-field variance.

The solid line in the figure shows a fit to the LF by the usual Schechter function \cite{sc76},
\begin{equation}
\Phi (M)\propto e^{-10^{0.4(M^*-M)}}10^{-0.4(\alpha +1)M},
\end{equation}
where $M^*$ is the characteristic magnitude, representing the turnoff at the bright end of the LF profile, and $\alpha$ is the slope beyond the up-turn at the faint end of the LF. We find $\alpha~=~-1.42~\pm~0.12$, and a characteristic apparent $R$--band magnitude  $m^*_R~=~20.8~\pm~0.6$.  The errors in the Schechter function parameters are calculated from Monte Carlo simulations, where we add, or subtract, a random number of galaxies to each magnitude bin according to the uncertainties. The quoted errors are the 1$\sigma$ deviations in the parameters from one hundred Monte Carlo realisations. The LF for the background field at the cluster redshift has $m^*_R~=~20.6~\pm~0.2$, and $\alpha~=~-0.90~\pm~0.08$. This is similar to the slope for the total LF for the field, $\alpha~\sim~-1$, found by e.g. Loveday et al. \shortcite{lo92} and Marzke, Huchra \& Geller \shortcite{ma94}.

The determination of $M^*$ is based primarily on the brightest galaxies. Here statistics are poor, and the error in $M^*$ significant. For example, excluding the two brightest bins, with only three galaxies, increases $M^*$ by $\sim~0.6$ mag. There is also a strong coupling between $M^*$ and $\alpha$, making the faint-end slope dependent on the derived value of the turnoff magnitude. In Section 5.1 we show that a Gaussian plus Schechter function fit for the cluster LF gives a better representation of the underlying population. 
To avoid the coupling between $M^*$ and $\alpha$ we also calculate the faint-end slope by fitting the five faintest magnitude bins, i.e. $\sim$ 2.5 mag brighter than the limiting magnitude, to a straight line 
(as proposed by Trentham 1998). The line has the form
\begin{equation}
\Phi_f (M)\propto 10^{-0.4(\alpha_f +1)M},
\end{equation}
where the index $f$ denotes that only the faint magnitude bins are used. For Cl1601+42 this yields $\alpha_f~=~-1.38~\pm~0.07$. For the background field we find $\alpha_f~=~-1.17~\pm~0.10$.

In order to compare our results to the slope for other more nearby clusters, we must add K--corrections to calculate the absolute rest frame magnitudes in different bands. A major advantage of the photometric method is that individual K--corrections for each galaxy are known, based on the best-fitting template. In the standard one-filter subtraction method, a single value for the K--correction usually has to be {\it assumed}, due to the difficulty of determining the galaxy type of the faint objects from single band observations (e.g. Driver et al. 1998a; N\"{a}slund et al. 2000).

\begin{figure*}
\begin{minipage}{140mm}
\psfig{figure=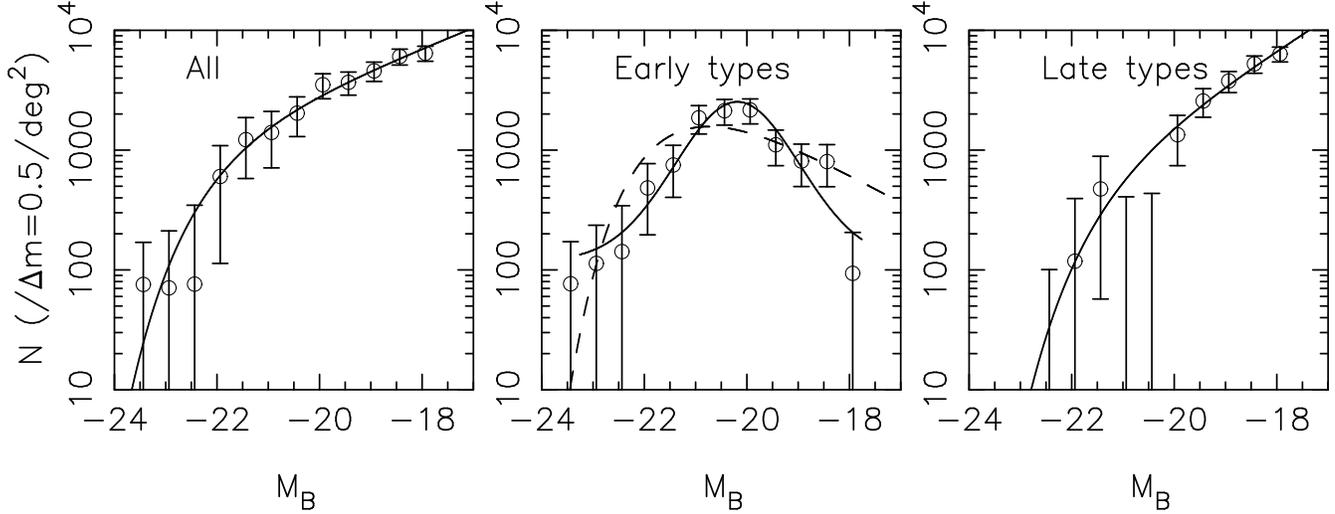,height=6.8cm,angle=-90}
\caption{Luminosity functions in the {\it rest frame} $B$--band for Cl1601+42. The left panel shows the total LF, whereas the middle and right panel shows the LFs where the galaxies are divided into early-types and late-types. The total and late-type LFs are fitted to a Schechter function. The solid line in the middle panel shows a Gaussian fit, while the dashed line is based on the Schechter function. Note the better fit of the Gaussian.}
\label{Figure5}
\end{minipage}
\end{figure*}

In the left panel of Fig. \ref{Figure5} we plot the resulting {\it rest frame} $B$--band LF. For a constant K--correction this should be similar to the observed $R$--band LF. Because of the variation in the individual K--corrections of the different galaxy types there is, however, a small redistribution of the galaxies between the bins, leading to a slight steepening of the faint-end slope. The turnoff magnitudes, faint-end slopes and absolute magnitude limits for the rest frame LFs in all bands are listed in Table \ref{Table2}. The absolute magnitude limits for the different filters approximately correspond to $R~=~25$, and are calculated from the average of the individual K--corrections for the galaxies.
\begin{table*}
\begin{minipage}{140mm}
\caption{Parameters for the LF in the rest frame for different filter bands. $M^*$ and $\alpha$ are the characteristic magnitude and faint-end slope of the Schechter function, while $\sigma_{M^*}$ and $\sigma_{\alpha}$ are the errors in these parameters. $\alpha_f$ is the slope of the straight line fit to the five faintest magnitude bins, and $\sigma_{\alpha_f}$ is the error in this quantity.}
\begin{tabular}{cccccccc}
\bf Filter &\bf Magnitude limit & \bf $M^*$ & $\sigma_{M^*}$ & \bf $\alpha$ & $\sigma_{\alpha}$ & $\alpha_f$ & $\sigma_{\alpha_f}$\\ 
& & & & & & & \\
$U$ & -17.9& -21.8& 0.9& -1.46 & 0.25 & -1.49 & 0.12\\ 
$B$ & -17.6& -22.3& 0.5& -1.37 & 0.07 & -1.41 & 0.09\\ 
$V$ & -18.4& -22.5& 0.8& -1.40 & 0.14 & -1.44 & 0.09\\ 
$R$ & -18.9& -24.0& 0.9& -1.48 & 0.11 & -1.49 & 0.09\\
$I$ & -19.5& -24.8& 0.9& -1.41 & 0.11 & -1.45 & 0.10\\ 
\end{tabular}
\label{Table2}
\end{minipage}
\end{table*}

\subsubsection{The cluster LF for different populations}
To investigate the LF of the different galaxy types, we divide the cluster galaxies into early-type and late-type subsamples, as described in Section 3.5. It is important to note that this refers to a spectroscopic type, not morphological. Field galaxies within the cluster redshift range are subtracted with a distribution of galaxy types determined from the EIS data. The late-type fractions in the cluster to different limiting magnitudes are listed in Table \ref{Table3}. Early-types dominate at bright magnitudes, whereas late-types dominate at magnitudes somewhat fainter than $R~\sim~23$. 

The distinction between a blue population that dominates at faint magnitudes, and a population of bright red early-type galaxies is illustrated in Fig. \ref{Figure5}, where we plot the total rest frame $B$--band LF, as well as the LFs for the two populations separately. The faint-end slope and characteristic magnitude for the late-type subsample are $\alpha~=~-1.68~\pm~0.11$ and $M^*_B~=~-21.4~\pm~0.6$. For the early-type subsample we find $\alpha~=~-0.41~\pm~0.20$ and $M^*_B~=~-21.3~\pm~0.4$. The low statistics at $M_B~\lsim~-20.5$, of the late-type LF makes this part, as well as $M^*_B$, uncertain. The straight line fit for $M_B~\gsim~-20.1$ results in $\alpha_f~=-1.80\pm~0.14$, and $\alpha_f~=0.17\pm~0.42$, for the late-type and early-type subsamples, respectively.

It is clear that the faint-end slope of the late-type LF is steep, and that this blue population also dominates the faint end of the common LF. The red population shows an opposite behaviour at faint magnitudes, and is better fitted by a Gaussian. The reduced chi-square of the Gaussian fit to the early-type LF is $\chi^2/\nu~\sim~0.54$, while the Schechter function fit yields $\chi^2/\nu~\sim~1.0$, clearly showing that a Gaussian gives a better representation of the LF.
\begin{table}
\caption{The fraction of the total number of galaxies that are late-type spirals or irregulars at different limiting magnitudes in the $R$--band.}
\begin{tabular}{lcccc}
\bf Magnitude, $m_R$ & \bf $<$ 23 & \bf 23--24 & \bf 24--25\\ 
& & & \\
Late-type fraction & 0.41 & 0.89 & 0.96 \\ 
\end{tabular}
\label{Table3}
\end{table}
%\subsubsection{The cluster LF from the standard background subtraction method}
\subsubsection{The standard background subtraction method}
As a comparison, we calculate the LF using the standard background subtraction method in a single filter. As background we use our $R$--band image of a blank field, and subtract this from the cluster image, correcting for the slightly different field sizes. With this method we reach $R~=~26$, i.e., about one magnitude fainter than for the photometric redshifts. For the two faintest magnitude bins we have corrected for incompleteness as described in Section 2.5. This results in $552~\pm~42$ cluster galaxies to $R~=~26$, while the number of galaxies to $R~=~25$ is $373~\pm~27$, where the errors are statistical only. Including field-to-field variance in the background counts results in $373~\pm~120$ at $R~=~25$. To compare this with the number found by the photometric method, we correct the latter for the galaxies lost due to the 1.5$\sigma$ cut in redshift range, and include field-to-field variance. This results in $309~\pm~31$ at $R~=~25$, which is within the errors found by the subtraction method.

\begin{figure}
\psfig{figure=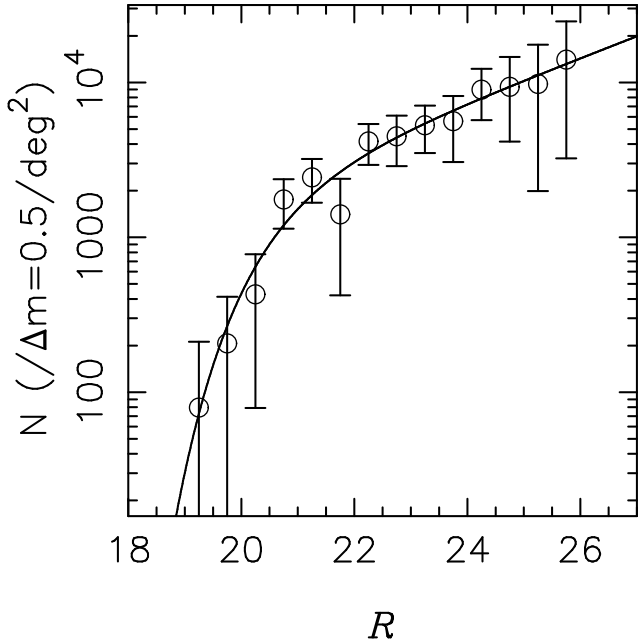,height=8.5cm,angle=-90}
\caption{Luminosity function in the $R$--band for Cl1601+42 calculated by the standard method of subtracting a background field from the cluster image. Error-bars include Poissonian uncertainties and field-to-field variance.}
\label{Figure6}
\end{figure}

The LF to $R~=~26$ using the subtraction method is shown in Fig. \ref{Figure6}. A Schecter fit yields a faint-end slope $\alpha~\sim~-1.38~\pm~0.12$, and the turnoff magnitude is $m^*_R~=~20.6~\pm~0.3$, close to the values derived using photometric redshifts. The faint end fit, as defined by Eq. (3), for both $R_{lim}~=~25$ and $R_{lim}~=~26$ results in $\alpha_f~\sim~-1.44~\pm~0.11$ which shows that there is no dramatic change at the faint end when going slightly deeper in magnitude. Comparing the error-bars in Figure \ref{Figure4} and Figure \ref{Figure6} clearly illustrates the gain with the photometric method, especially at faint magnitudes. The larger error-bars in Figure \ref{Figure4} around $R~\sim$ 21 are due to the smaller statistics in these bins for the photometric method, probably caused by the use of different background fields in the two methods.

The result also indicates that the subtraction method works reasonably well to redshifts $z~\sim~0.55$, even though the number of cluster galaxies vary somewhat between the methods. This result may seem contrary to Driver et al. \shortcite{dr98a}, who from simulations find that the method is unreliable for clusters at $z~>~0.3$. Their conclusions are, however, valid for a specific observational set-up. We reach a completeness limit $\sim~1.5$ mag fainter than the limit used by Driver et al., which should increase the reliability of the subtraction method to a higher redshift.
\subsubsection{Gravitational lensing and crowding in the subtraction method}
A possible source of uncertainty in the background subtraction method comes from gravitational lensing. Background galaxies in the cluster image can have their magnitudes amplified by the cluster, leading to an excess of background galaxies in the cluster image, compared to the cluster free image used for subtraction. For the rich cluster A963, which show signs of arcs, Driver et al. \shortcite{dr94} estimate that the increase in background counts is $\sim$ 10 per cent. When investigating the Coma cluster, Bernstein et al. \shortcite{ber95} argue that this increase will be cancelled by the reduction of the solid angle being viewed behind the lensing cluster, making no correction necessary. Considering this, together with the fact this Cl1601+42 is both intermediate in richness and concentration and has no detected arcs, we find that gravitational lensing should not affect our results. 
Effects from gravitational lensing should not enter into the photometric method, since we do not subtract galaxies behind the cluster, but instead reject them using the redshift information.

Another factor to account for in the subtraction method is crowding, i.e. overlapping object. Especially faint objects, with small areas, can be hidden behind brighter galaxies and stars. Since there are in total more objects in the cluster image than in the background image, this can lead to a higher fractional loss of faint galaxies in the cluster image, resulting in a LF with too few faint objects. Using similar arguments as in N\"{a}slund et al. \shortcite{na00}, we estimate the loss by adding the total area covered by objects in the cluster image and the background image, respectively. We find that the correction in the number of faint galaxies in the cluster LF is less than two per cent. Since this correction factor is much less than statistical errors, as well as field-to-field variations, we do not include any corrections due to crowding in our results.

For the photometric method, only foreground galaxies contribute to the crowding. The effect should be of equal importance in the cluster field and the field used for subtraction, and therefore cancel.

\subsection{Radial distribution}
\begin{figure*}
\begin{minipage}{140mm}
\psfig{figure=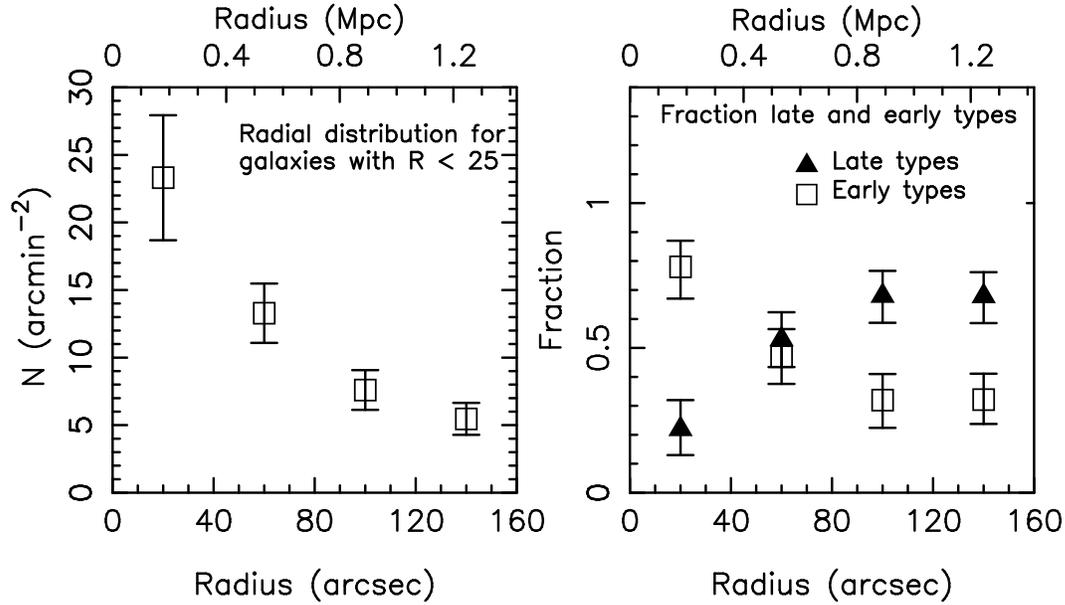,height=8.0cm,angle=-90}
\caption{The left panel shows the radial distribution of the cluster galaxies in Cl1601+42. Background subtraction within the cluster redshift range has been applied. The right panel shows the fraction of early-type and late-type galaxies in the cluster as a function of radius. In absolute magnitudes $R~<~25$ corresponds to $M_B~<~-17.6$ in the rest frame.}
\label{Figure7}
\end{minipage}
\end{figure*}

Using the cluster galaxies selected by photometric redshifts, we plot the radial distribution of the cluster members with $R$ $<$ 25 in Fig. \ref{Figure7}. The left panel shows the projected surface density, and the right panel shows the fraction of late-type and early-type galaxies as a function of radius.

The most interesting result is that the cluster outside of $\sim~0.7$ Mpc shows a roughly constant fraction of late-type galaxies, $\sim$ 70 per cent of the total. Inside this radius the fraction of early-type galaxies increases rapidly to $\sim$ 80 per cent at the centre. Although, this effect is present also in low redshift clusters (e.g. Whitmore, Gilmore \& Jones 1993), our result shows that this population gradient is set up already at $z \sim$ 0.54. In fact, the gradient is even steeper than in local clusters.

\begin{figure*}
\begin{minipage}{140mm}
\psfig{figure=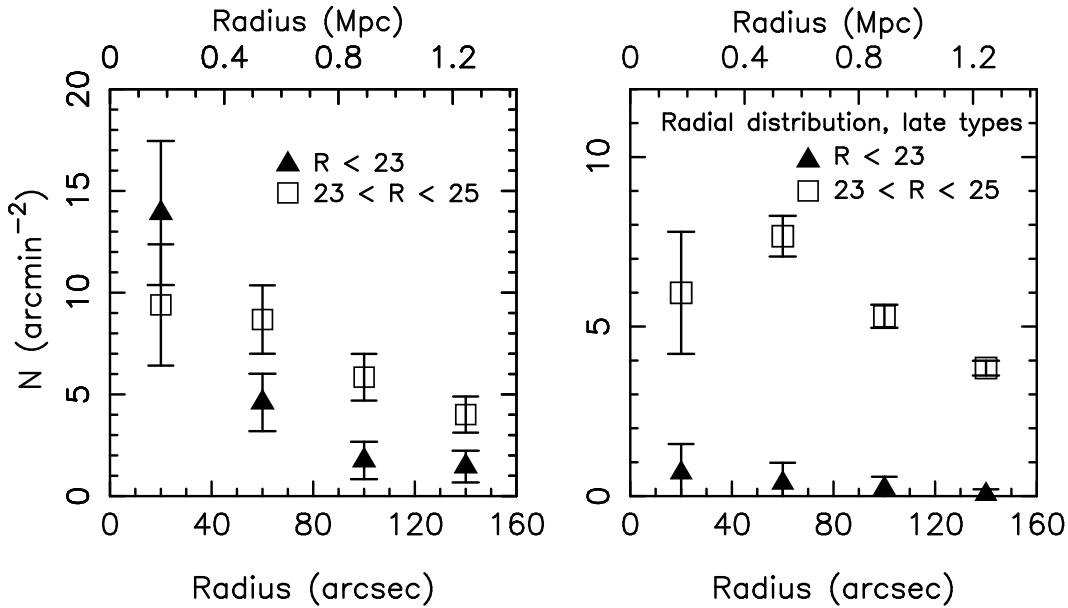,height=8.0cm,angle=-90}
\caption{The left panel shows the radial distribution of bright ($R~<~23$, $M_B~\lsim~-19.6$) and faint ($23~<~R~<~25$, $-19.6~\lsim~M_B~\lsim~-17.6$) galaxies in Cl1601+42. The right panel shows the radial distribution of bright and faint late-type galaxies.}
\label{Figure8}
\end{minipage}
\end{figure*}

In Fig. \ref{Figure8} we plot the radial distribution for bright ($R~<~23$) and faint ($23~<~R~<~25$) galaxies separately. While the bright galaxies show a steep cluster-centric distribution, the faint galaxies have a slower increase, which flattens at the centre, consistent with a homogeneous distribution of faint galaxies in the cluster. The figure also shows the radial distribution separately for bright and faint late-type galaxies. We discuss the radial distribution further in Section 5.2.
\subsection{The Butcher--Oemler effect}
BO84 showed that the fraction of blue galaxies, $f_B$, in cores of rich compact clusters increases dramatically from $f_B~\sim~0.03~\pm~0.01$ for clusters with $z~\lsim~0.1$, to $f_B~\sim~0.25$ at $z~\sim~0.5$.

To determine $f_B$ in Cl1601+42 we adopt the definition in BO84. First, galaxies brighter than $M_V~=~-20$ within a circular radius, $R_{30}$, containing 30 per cent of the cluster galaxies are selected. From this sample $f_B$ is defined as the fraction of these galaxies with rest frame colours $B-V$ at least 0.2 mag bluer than the linear locus of the colour-magnitude (CM) relation. In order to compare our results with those of BO84 we use in this section their adopted cosmology, i.e $H_0~=~50$~km~s$^{-1}$~Mpc$^{-1}$ and $q_0~=~0.1$.

We find $R_{30}~\sim~1.2$ arcmin, or $\sim~0.6$ Mpc
for the cosmology above. This is somewhat smaller than the typical
radius of $R_{30}~\sim~1$ Mpc given by Kodama \& Bower
\shortcite{ko01} for a sample of rich clusters at
$0.2~\lsim~z~\lsim~0.4$, but similar to the radius $R_{30}~\sim~0.5$ Mpc used by van Dokkum et al. \shortcite{va00} for the cluster MS1054-03 at $z~=~0.83$. Our value of  $R_{30}$ is also similar in angular size, $\sim~1$ arcmin,
to that used by BO84 for their highest redshift clusters. The blue fraction we find is $f_B~=~0.15~\pm~0.06$. 

This value for Cl1601+42 is significantly higher than the local value $f_B~\sim~0.03$, but lower that the average value $f_B~\sim~0.27$ at $z~=~0.54$, taken from the fit presented by BO84. 
Note, however, that BO84 did not use K--corrections that depend on the spectral type of the galaxies \cite{ko01}. Including this, $f_B$ decreases by $~\sim~0.04$, and the K-corrected average BO value at $z = 0.54$ would therefore be $f_B \sim 0.23$. Even after this correction, we find that a significant difference between the blue fraction in Cl1601+42 and the BO84 sample remains. 

Finally, a possible concern here are systematic effects, which can discriminate against selecting blue late-type galaxies, and therefore lead to an underestimate of $f_B$. From the MC simulations described in Section 3.3, as well as comparing with available spectroscopic redshifts, we find that systematic effect can cause a relatively higher loss of blue galaxies when selecting cluster galaxies by using a fixed redshift interval. This may lead to an underestimate of the blue fraction by $\Delta f_B~\sim~0.01$. The small effect here is due to the relatively bright limiting magnitude used to select the BO-galaxies, i.e. $M_V~<~-20$ (corresponding to $m_R~\lsim~23.5$). Here systematic effects are not expected to affect the photometric redshifts more than marginally. However, when we in Section 5.3 determine the blue fraction to fainter magnitudes, the underestimate may reach $\Delta f_B~\sim~0.05$.

\subsection{Comparison with Hubble types from HST}
Cl1601+42 was observed with WFPC--2 on HST during Cycle 4 with the F702W filter. The total exposure was 16800 s \cite{sm97}. The image covers the central 5 square arcmin of the cluster, and detailed morphological information are presented by Smail et al. for all 145 objects with $R_{\rm{F702W}}<~23.5$. Five of these are classified as stars, whereas no classification was possible for a further six objects. Of the five objects classified as stars we get the same identification for three, while the other two are found in the lowest redshift bin, consistent with these also being stars. Of the six objects without HST classification we identify three as stars, one as possible star and two as galaxies. This leaves 134 galaxies with classified morphology. We are able to determine photometric redshifts for 129 of these objects.

Of the 68 galaxies classified as early-type galaxies by the photometric redshift method, $\sim~62$ per cent have morphologies classified as E to Sa, and the rest are Sab to Irr. Conversely, of the 61 galaxies photometrically classified as late-type galaxies, $\sim~72$ per cent are morphologically classified as Sab to Irr. Even though there may be some uncertainty in the visual classification into Hubble types (Smail et al. set $R_{F702W}<~23.5$ as the limit where no more than 20 per cent of the galaxies differ by more than one Hubble type between two different classifiers), the main reason for this relatively low correspondence is the lack of a direct relation between spectroscopic and morphological types, as also shown by e.g., Dressler et al. \shortcite{dr99} and Poggianti et al. \shortcite{po99}.

\begin{figure}
\psfig{figure=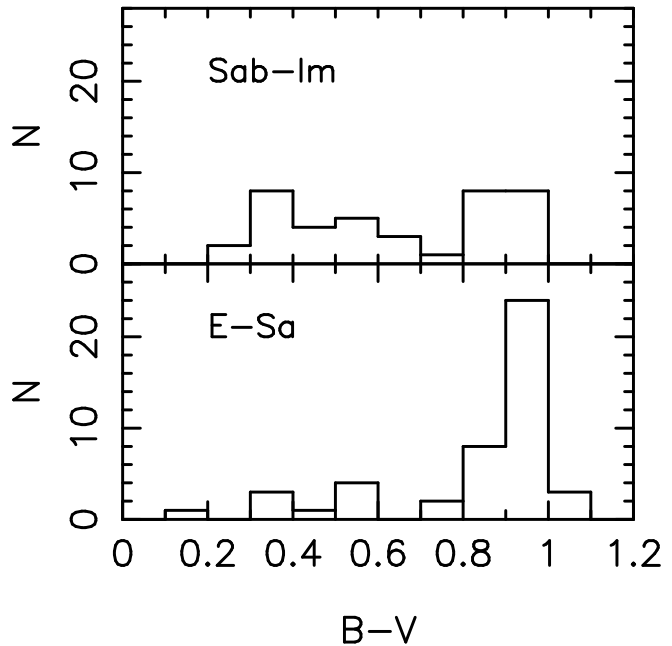,height=8.5cm,angle=-90}
\caption{Rest frame $B-V$ colour distribution of early-type galaxies (E--Sa) and late-type galaxies (Sab--Im) brighter than $R_{F702W}~=~23.5$ in Cl1601+42. Morphologies are as determined by Smail et al. (1997).}
\label{Figure9}
\end{figure}

This can be seen in Fig. \ref{Figure9}, which shows the rest frame $B-V$ colour for the galaxies within the redshift range of the cluster, divided into two broad morphological classes. The lower panel shows the colours of the galaxies with early Hubble types (E--Sa), and the upper panel later types, as determined from the HST data by Smail et al. Most early-type galaxies are red and have a small dispersion in colour. There is, however, a tail of galaxies with relatively blue colours. The distribution of later types have a broader distribution, but are on average $\sim~0.2$ mag bluer than the earlier types. The figure clearly shows that there is relatively large population of objects, morphologically classified as late-type galaxies, that have red colours.
\section{Discussion}
\subsection{Evolution of the cluster luminosity function}
Luminosity functions for galaxies fainter than $M_B~\sim~-18$ have previously been determined for clusters to $z~\sim~0.3$. Comparing the LFs for Coma at $z~=~0.023$ \cite{tr98}, A2554 at $z~=~0.106$ (Smith et al. 1997), A963 at $z~=~0.206$ \cite{dr94} and MS2255.7+2039 at $z~=~0.288$ (N\"{a}slund et al. 2000) reveals that both nearby and distant clusters show a steep faint-end slope ($\alpha~\lsim~-1.4$), indicating no qualitative difference with redshift. Our results for Cl1601+42 show that this conclusion is true up to $z~\sim~0.5$.

For the highest redshift cluster with a deep LF, MS2255.7+2039 at $z~=~0.288$, N\"{a}slund et al. \shortcite{na00} derive $M_R^*~=~-24.2$ (in our cosmology) and $\alpha~\simeq~-1.43$, when fitting their data to a single Schechter function. This agrees well with our results in Table \ref{Table2}.

The increased fraction of faint galaxies in the outer part of Cl1601+42, shown in the left panel of Fig. \ref{Figure8}, has the consequence that the faint-end slope of the LF should depend on radius. Consequently, we find that the faint-end slope for galaxies within a radius of 80 arcmin (0.7 Mpc) is $\alpha_f~\sim~-1.27$, while between 80--160 arcmin (0.7--1.4 Mpc) $\alpha_f~\sim~-1.52$. When comparing the LF's between clusters, it is therefore important to remember that the slope of the LF in the core region can be much shallower than for the cluster as a whole. 

A steepening of the slope with radius is consistent with studies of local clusters. Adami et al. \shortcite{ad00} e.g., find a faint-end slope $\alpha~\sim~-1.0$ for the core region of Coma, while for the whole cluster Trentham \shortcite{tr98} finds a considerably steeper faint-end slope, $\alpha~\sim~-1.7$. Also, five of seven clusters at $z~\sim~0.15$ examined by Driver, Couch \& Phillipps \shortcite{dr98b} show this trend. The faint-end slopes for these seven clusters vary between -0.9 and -1.4, which is more shallow than the slopes for the clusters quoted in the beginning of this section. The flatter slope compared to Cl1601+42 may be understood as a result of the effect above, since the clusters in the sample of Driver et al. only cover a radius half the linear size of our cluster. 
 
\begin{figure}
\psfig{figure=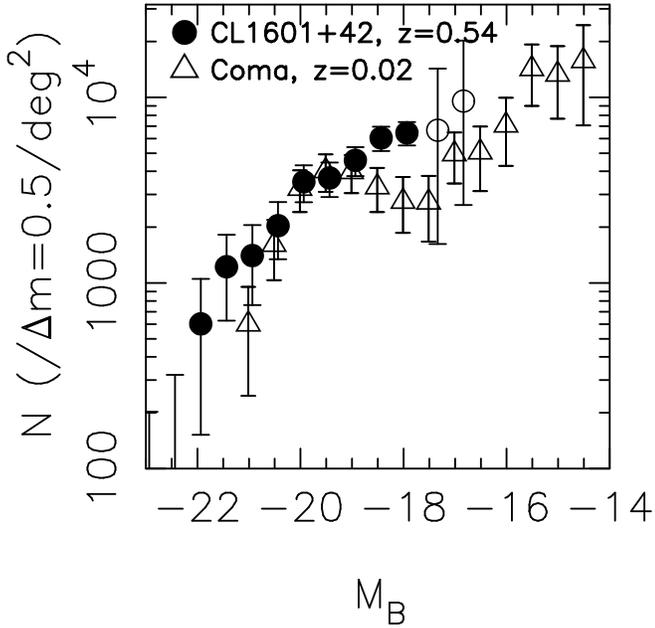,height=8.3cm,angle=-90}
\caption{$B$--band LFs for Cl1601+42 and Coma (from Trentham 1998). Note the different shapes of the LF at intermediate magnitudes $-20.5~\lsim~M_B~\lsim~-18.0$. At the faint end of Cl1601+42 we have added two data points derived from the single-filter subtraction method (shown as open circles). Coma is normalised to Cl1601+42 at $M_B~\sim~-20$.}
\label{Figure10}
\end{figure}

It is interesting to compare the rest frame $B$--band LFs of Cl1601+42 and the Coma cluster (Trentham 1998), shown in Fig. \ref{Figure10}. Coma is here normalised to Cl1601+42 at $M_B~\sim~-20$. Note that we use similar methods as Trentham to correct the magnitudes at faint levels (see Section 2.5). Also, both investigations use a 3$\sigma$ detection and similar surface brightness limits. We have for Cl1601+42 added two data points derived from the one colour subtraction method (open circles). These points are in agreement with a continuation of the steep slope found at intermediate magnitudes. The figure shows that at intermediate magnitudes, $-20.5~\lsim~M_B~\lsim~-18.0$, Coma has a Gaussian LF peaking at $M_B~\sim~-19.5$, before the faint galaxies start to dominate at $M_B~\gsim~-18.0$, and the slope of the LF turns steep. However, for Cl1601+42 the steep part sets in already at $M_B~\sim~-20$, and no plateau is present, as in Coma. 

If the apparent lack of a 'plateau' in Cl1601+42 is a general property at this redshift, it could indicate a general brightening of the faintest cluster members, because of more active star formation, with increasing redshift. This brightening would shift the faint, steep end towards brighter magnitudes at higher redshift, and therefore decrease the extent of the 'plateau'-range, while preserving the slope. This is consistent with Coma having $\alpha_f~\sim~-1.5$ at $M_B~\gsim~-17$, and Cl1601+42 $\alpha_f~\sim~-1.4$ at $M_B~\gsim~-19$.
 
The 'plateau' between the turnoff at bright magnitudes, and the up-turn at faint, is present in both MS2255 at $z~=~0.288$ and A963 at $z~=~0.206$. This indicates that the star formation activity of the faint galaxies indicated by Cl1601+42 at $z~=~0.54$, is not as important in the two clusters at lower redshift. 

There is some controversy regarding the colour dependence of the
faint-end slope of the LF. De Propris et al. \shortcite{de95}
determine the LF for four Abell clusters in both $B$ and $I$. Although
the slopes are very steep ($\alpha~<~-2$), there is no difference
between the two colours. These results are in contrast to the results
of Wilson et al. (1997), who determine the LF in $B$ and $I$ for A665
and A1689, both at $z~=~0.18$, and find $\alpha~\sim~-2$ for the
$V$--band and $\alpha~\sim~-1.2$ for the $I$--band. Comparing the
slope of the LF's of the different colours for Cl1601+42 (Table
\ref{Table2}), there is no significant trend when going from red to
blue filters. This is consistent with the dominance of the late-type population at faint magnitudes. If a majority of the galaxies have similar colours, i.e. K-corrections, then one expects a horizontal shift of the faint-end LF between the different bands, while the steepness remains unchanged.

To investigate if there is a trend at even shorter wavelengths in Cl1601+42, we calculate $\alpha_f$ at rest frame $UV(2850$ \AA) and $UV(2350$ \AA), corresponding to the observed $B$-- and $U$--bands, and find $\alpha_{f2850}~=~-1.71~\pm~0.20$, and $\alpha_{f2350}~=~-2.05~\pm~0.20$, for $M_{UV(2850)}~\gsim$ -19.0 and $M_{UV(2350)}~\gsim$ -19.1, respectively. 

While the optical bands do not show any trend in the faint-end slope, the steepening of the slope in the $UV$-bands suggests that the faintest cluster galaxies have an increasing fraction of young, $UV$ bright stars.

In a recent paper Valotto, Moore \& Lambas \shortcite{val01} discuss systematic effects when determining the LF for clusters selected in two dimensions, such as the Abell and EDCC (Edinburgh--Durham Cluster Catalogue, Lumsden et al. 1992) catalogues. From N-body simulations Valotto et. al. find a bias towards selecting clusters which are contaminated by background clusters, or groups, projected on to the cluster. The over-density in the background, along the cluster line-of-sight, will result in an over-abundance of faint galaxies, which remains after the subtraction of a cluster-free background field. For a composite LF, calculated from simulations where the clusters are selected in two dimensions, similar to the Abell clusters, Valotto et al. find 
%$\alpha~\sim~-1.4$, when using statistical background subtraction, which is steeper than the input LF, for which they assume $\alpha~=~-0.97$. The 
that the selection process therefore can introduce an artificial steepening of the LF.

By selecting cluster members with photometric redshifts, the background contamination should be minimized in our sample. The steep faint-end slope that we find should therefore not be much affected by the selection effects described by Valotto et al. Furthermore, the similar slope we derive from the background subtraction method indicates that projection effects are not important for Cl1601+42.
\subsubsection{Evolution of the LF for different cluster populations}
From deep studies of both near and intermediate redshift clusters, it is clear that a single Schechter function gives a rather poor representation of the LF (e.g. Driver et al. 1994; Wilson et al. 1997; N\"{a}slund et al. 2000). For the Virgo cluster Binggeli et al. \shortcite{bi88} find that ellipticals with $M_B~\lsim~-18$, as well as the different types of spiral galaxies, are better represented by Gaussian LFs. The mean magnitude is for the spirals shifted towards fainter magnitudes for the later types. 
The sum of all types of spirals can also be well represented by a Gaussian LF.  
%Summing up the different spiral types also results in a Gaussian LF. 

At magnitudes M$_B~\gsim~-18$ dwarf ellipticals (dE) and irregulars dominate. The up-turn of the LF at the faintest magnitudes is mostly due to the dE population, where a Schechter function is a better representation of the LF. The total LF should therefore rise as a Gaussian at bright magnitudes and then flatten to a 'plateau' at intermediate magnitudes, before it again turns steep at the faint end.

The conclusions above may seem to contradict other studies showing a flat LF. Garilli, Maccagni \& Andreon \shortcite{ga99} find $\alpha~\sim~-0.9$ for a composite LF calculated from 65 clusters at $0.05~<~z~<~0.25$, while Paolillo et al. \shortcite{pa01} find $\alpha~\sim~-1.1$ for 39 clusters at $0.058~<~z~<~0.28$, when fitting their data to a single Schechter function. Paolillo et al. do, however, point out that the LF below M$~\sim~-19.5$ in these clusters, which probes the dwarf population, is systematically above the Schechter function fit, indicating a upturn of the faint-end slope at these magnitudes. The flat slopes found in these investigations are therefore not in conflict with the steeper slopes found in deeper studies. 

Above we argued that the lack of a plateau in the LF of Cl1601+42, as compared to clusters at lower redshift, can be explained by a uniform shift in magnitude for the fainter galaxies. Because these galaxies are predominantly blue (Fig. \ref{Figure5}), the proposed shift can be interpreted as a result of a higher star formation level in this population of galaxies compared to the bright population. 
For cluster galaxies at $z$ = 0.24, which have had their star formation truncated, Smail et al. (1998) show that late types galaxies both fade more and become redder than early type galaxies. This is consistent with the shift of the late type population in Cl1601+42. The total reddinging and fading between $z$ = 0.24 and $z$ = 0 is, however, fairly modest in these models. A more substantial fading is expected if dynamical processes are involved.

Wilson et al. \shortcite{wi97} argue that dwarf irregulars in clusters should fade rapidly, either due to a trunction process, such as ram pressure stripping, or by disruption due to tidal interactions. They estimate the fading since $z~\sim~0.2$ to $\gsim~3$ mag. Such a fading would make part of the progenitors of todays dwarfs with $M_B~\lsim$ -15 detectable in Cl1601+42, and could therefore contribute to the faint-end slope. Further, simulations of 'galaxy harassment' \cite{mo98} show that high redshift ($z~\sim~0.4$) cluster galaxies of type Sc or later can be turned into low surface brightness spheroidal systems in todays local clusters. The fading of these 'harassed' galaxies should be $\sim~2$ mag. The dwarf ellipticals present at $z~\sim~0.5$ should, on the other hand, only fade marginally by $\lsim$ 0.5 mag. 
Whatever the mechanism, it seems that at least a fraction of the galaxies contributing to the steep slope at the faint end in local clusters have faded considerably compared to earlier epochs. The gradual disappearance of the plateau at high redshift could therefore be explained as a fading of the faint, late-type population, relative to the bright population. 

This scenario requires that the number of recently accreted galaxies is higher at $z~\sim~0.5$ than today, i.e. that there is an increase in the infall rate with redshift. As shown by the Press--Schechter formalism, N-body simulations, as well as observations, this is indeed the case \cite{bo91,ka95,di01,el01,ko01}.
A bluening with redshift, i.e. an increased star formation rate, of the infalling field population can also contribute to the observed effect. Diaferio et al. \shortcite{di01} show that there is a significant increase in star formation rate in field galaxies over the redshift range $0.18~<~z~<~0.55$. Therefore, the brightening is likely to be due to a combination of the above two effects. 
The brightening of the blue cluster population also relates to the BO effect, which shows an increase of star formation in a large fraction of cluster galaxies at higher redshifts. We return to this in Section 5.3.

We now focus on the evolution of the bright cluster population. Smail et al. \shortcite{sm97} calculate composite LFs for ten clusters in the MORPHS sample divided into three redshift bins centred on $z~=~0.38$, $z~=~0.43$ and $z~=~0.54$. (The last bin includes Cl1601+42.) The galaxies are here divided into ellipticals and spirals (Sa and later), based on visual classification into morphological types. The limiting magnitude for this sample is, however, too bright to detect the dwarf population, or for determining the faint-end slope. Instead, when calculating the turnoff magnitude, Smail et al. fix the faint-end slope at $\alpha~=~-1.25$ for the ellipticals, and at $\alpha~=~-1.0$ for the spirals. They find that there is a slow brightening of the elliptical population from $z~=~0$ to $z~=~0.54$ by $\Delta M~\sim~-0.3$. The spiral population does, however, not show any clear evolution with redshift. A change of cosmology affects these results somewhat. Adopting $\Omega_M~=~0.3$ and $\Omega_{\Lambda}~=~0.7$, instead of $\Omega_M~=~1.0$ as in Smail et al., increases the brightening of the ellipticals to $\Delta M~\sim~-0.7$. There will in this case also be a marginal brightening for the spiral galaxies. A somewhat larger evolutionary trend is found by Kodama \& Bower \shortcite{ko01}, who estimate that the bright spiral population have faded by about one magnitude since $z~\sim~0.4$.
The fading of the bright population is therefore smaller than of the faint population, which is in turn consistent with the disappearance of the plateau.

\subsection{The radial population gradient}
As we have just discussed, galaxies are in a hierarchical scenario accreted from the field on to the clusters (e.g., Diaferio et al. 2001). When field galaxies, which are predominately spirals with ongoing star formation, are accreted, star formation is halted after a possible burst of star formation. The quenching can either be abrupt, or there could be a more gradual decline in star formation (e.g. Barger et al. 1996, Poggianti et al. 1999, Balogh et al. 1999, Balogh, Navarro \& Morris 2000, Diaferio et al. 2001). The decrease in star formation should set up a radial gradient in the clusters, where blue galaxies are expected to be more numerous in the outer parts, as is confirmed by observations and simulations (Diaferio et al. 2001). Also, the higher accretion rate at earlier epochs, as well as higher star formation rate in the general field, implies that the fraction of blue galaxies should be higher in high--$z$ clusters. There is therefore a connection between the shape of the LF and the existence of a radial colour gradient.

In the right panel of Fig. \ref{Figure8} we showed the radial distribution of late-type galaxies, divided into a bright subsample, with $R~<~23$, and a faint subsample with $23~<~R~<~25$. From the figure it is obvious that bright, blue galaxies are rare across the whole cluster. The flat, or at most slightly increasing, projected surface density is consistent with a decline in the spatial distribution towards the core, even though the small statistics do not make this significant.

The faint blue galaxies show a more dramatic trend. At radii $\gsim~0.4$ Mpc this is the most numerous population in the cluster. Here, the surface density is consistent with a volume density that is either uniform, or decreasing  with radius. The drop in surface density at the innermost point is consistent with an absense of faint blue galaxies at the core of the cluster. An explanantion may be provided by 'galaxy harassment', which should be increasingly efficient in the denser core region. 

The increasing infall rate at high redshifts, predicted by hierachical clustering, should result also in a change with redshift of the radial gradient of the fraction of early-type galaxies in the cluster. Ellingson et al. \shortcite{el01} show that galaxies with an old population dominate to larger radii in low redshift ($0.18~<z~<~0.30$) clusters, giving a flatter radial gradient, compared to high redshift ($0.30~<~z~<~0.55$) clusters. This indicates that high--$z$ clusters have a higher population of recently accreted, star forming galaxies at radii closer to the core, causing a steeper radial gradient. As the number of recently accreted star forming galaxies declines with time, the gradient becomes less steep. 

With similar magnitude limits and the same cosmology as in Ellingson et al., we find that the radial gradient of Cl1601+42 is at least as steep as the slope they find from their high redshift sample, and significantly steeper than that of the low redshift clusters.

\subsection{The radial and magnitude dependence of the Butcher-Oemler effect}
\begin{figure}
\psfig{figure=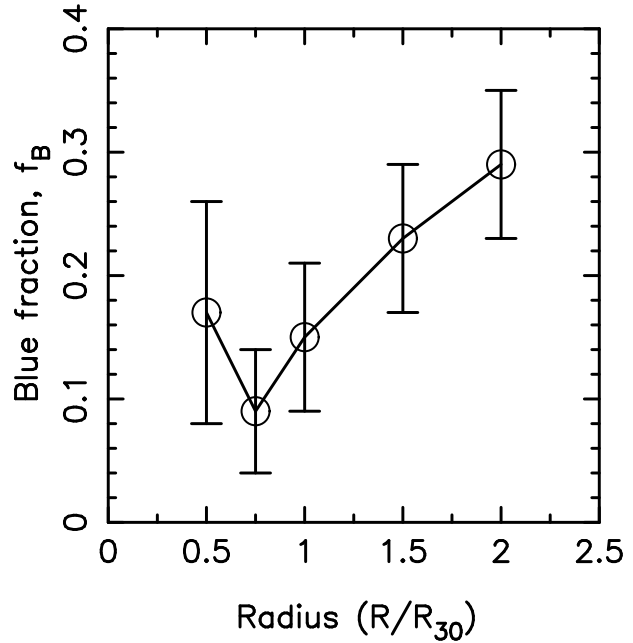,height=8.5cm,angle=-90}
\caption{The blue fraction, $f_B$, as a function of radius. $\rm{R_{30}}$ is the fiducial value used by BO84}
\label{Figure11}
\end{figure}

To check whether the blue fraction, $f_B$, decreases towards the core as a result of a decreasing star formation rate in the accreted field galaxies as they fall deeper into the cluster potential, we plot in Fig. \ref{Figure11} $f_B$ as function of radius (here given in terms of R$_{30}=0.6$ Mpc). From this figure it is clear that there is a trend for an increasing value of $f_B$ at larger radii. This is in agreement with both BO84 and Kodama \& Bower \shortcite{ko01}, and consistent with the infall scenario.
As is discussed by Kodama \& Bower, the change of $f_B$ with radius further illustrates the necessity of a correct determination of $R_{30}$ for a meaningful comparison of the blue fraction between clusters. 

It is also interesting to calculate the blue fraction as a function of the magnitude cut, as is done in Kodama \& Bower \shortcite{ko01} for their seven clusters at $0.23~<~z~<~0.43$. When varying the magnitude cut by $\pm~1$ mag from the fiducial value M$_V~=~-20$ they find that $f_B$ remains almost constant. Only for two of the most distant clusters is there a trend for $f_B$ to increase at fainter magnitudes, which Kodama \& Bower, however, attribute to increasing photometric errors with magnitude. Instead, they argue that $f_B$ should depend only weakly with magnitude.

\begin{figure}
\psfig{figure=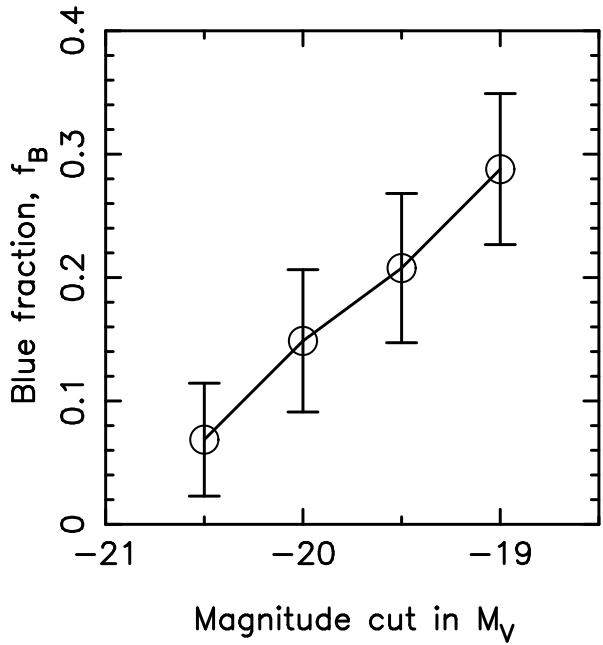,height=8.5cm,angle=-90}
\caption{The fraction of blue galaxies, $f_B$, for Cl1601+42 as a function of limiting magnitude. The fiducial value used by BO84 is $M_V$=-20.}
\label{Figure12}
\end{figure}

In Fig. \ref{Figure12} we plot $f_B$ as a function of the magnitude cut. We find a clear trend for an increasing value of $f_B$ as fainter galaxies are included, similar to the trend for two of the highest redshift clusters in Kodama \& Bower, but contradicting the general claim by Kodama \& Bower that the blue fraction should be constant with magnitude. As illustrated by Fig. \ref{Figure5}, the increasing value of $f_B$ at faint magnitudes in Cl1601+42 is a direct consequence of the different shapes of the red and blue LFs. The steepness of the blue LF causes the blue fraction to increase at fainter magnitude cuts, and decrease at bright. 

Because we are more than one magnitude brighter that our limit, the increase in photometric errors is $\lsim~0.03$ mag over the magnitude range where we calculate $f_B$. It is therefore unlikely that the strong trend of $f_B$ with magnitude in Cl1601+42 can be explained by increasing photometric errors at fainter magnitudes, as proposed by Kodama \& Bower for the clusters showing this trend in their investigation.

Note also that systematic effects when selecting the BO-galaxies may lead to an underestimate of the blue fraction in the faintest bin by $\Delta f_B~\sim~0.05$, as discussed in Section 4.4. This means that the trend shown in Figure \ref{Figure12} could be even stronger.

The dependence on the magnitude cut makes the calculated blue fraction dependent on the choice of cosmology. E.g., a constant cut in absolute magnitude, in the BO84 case $M_V~=~-20$, corresponds to a fainter apparent magnitude in a $\Lambda$--dominated cosmology than in a cosmology without cosmological constant. This should lead to a relatively higher blue fraction at increasing redshifts for the $\Lambda$ cosmology. Using $\Omega_M~=~0.3$ and $\Omega_{\Lambda}~=~0.7$, instead of the cosmology used by BO84 ($\Omega_M~=~0.2$, $\Omega_{\Lambda}~=~0$) changes the cut in apparent magnitude by $\Delta m~=~0.24$ at $z~=~0.54$. This increases the blue fraction from $f_B~=~0.15$ to $f_B~=~0.18$. 
%For clusters at lower redshift the increase in $f_B$ should be smaller
%, since the difference between cosmologies decreases at lower redshift. 
The difference between the extrapolated $f_B$ at $z~=~0.54$ from BO84, and the $f_B$ found for Cl1601+42 should therefore decrease in a $\Lambda$--cosmology. 

We finally notice that the blue fraction for the field, derived from the EIS data, is $f_B~\sim~0.40~\pm~0.07$ at the cluster redshift ($0.5~\lsim~z~\lsim~0.6$), and $f_B~\sim~0.39~\pm~0.08$ at lower redshift ($0.1~\lsim~z~\lsim~0.4$). This is close to the blue fraction $f_B~\sim~0.41~\pm~0.10$ calculated for the field by BO84. The similarity between the blue fraction in the low and intermediate redshift bins does not reflect the strong increase with redshift in the star formation in field galaxies found by Diaferio et al. \shortcite{di01}. This is somewhat surprising, even though the relatively large errors in $f_B$ do not rule out such trend. The fraction of late-type galaxies for the field is $\sim~0.7$. This fraction is consistent with the rough distribution of field galaxy types used by Smail et al. \shortcite{sm97}. 

\subsection{Metallicity versus star formation}
Colours of galaxies do not only reflect star formation, but are also dependent on metallicity, reflected in the CM relation (e.g. Faber 1977; Arimoto \& Yoshii 1987). In principle, a low metallicity could cause faint old, but blue, galaxies in our sample to be classified as star forming late-type galaxies. To estimate the importance of this effect for the blue population at $M_B \gsim -20$, we calculate the CM relation for the bright population with $M_B < -20$, and extrapolate this to fainter magnitudes. 

The slope of the $B-V$ vs. $V$ CM relation we find for the early-type galaxies in Cl1601+42 is $\Delta (B-V)/\Delta M_V~=-0.016$, for $M_V~<~-20.7$ ($M_B~\lsim~-20.0$). Over the full interval $-22~<~M_B~<~-18$ the mean bluening is then $\Delta (B-V) \lsim~0.07$. This is small compared to the mean off-set in colours between early-types and late-types, which is $\Delta (B-V) \sim~0.4$ at $M_B~\simeq~-18$. 

In addition, if part of the bluest old galaxies at faint magnitudes ($-20~<~M_B~<~-18$) are identified as late, star forming galaxies due to a low metallicity, the mean colour of the remaining early-type galaxies in this magnitude range will be redder than expected from an extrapolation of the CM relation. We do not see this effect, and therefore conclude that the bluening of early-type galaxies at faint magnitudes due to metallicity should not contribute more than marginally to the faint blue population we detect. The blue colours should therefore reflect the star formation.

Further support for this conclusion comes from Ferreras \& Silk \shortcite{fer00} who determine rest frame $NUV$ ($\sim$ 2000\AA) to $V$ colours of early-type galaxies in Cl0939+4713 at $z~=~0.42$. Comparing the observed colours to synthetic models with different star formation histories, metallicities and dust models, they find that the bluening at faint magnitudes is consistent with at least a fraction of the stars in the galaxies being young. Low metallicities alone are insufficient to explain the observed colours. (For further discussions on metallicity effects in cluster galaxies see e.g., Couch et al. 1998; Dressler et al. 1999; Poggianti et al. 1999.)

\section{Conclusions}
Using photometry in five broad-band filters we determine photometric redshifts for the galaxies in the field of Cl1601+42 at $z~\sim~0.54$. Cluster membership can therefore be assigned with a higher reliability for the individual galaxies, which, compared to statistical methods, significantly reduces the amount of background subtraction needed. As a by-product we also obtain information about the spectral properties of the galaxies, making assumptions about e.g. K--corrections unnecessary. Using this method, we calculate the luminosity function, the radial distribution of galaxies within the cluster, as well as the distribution of early-type and late-type galaxies. With this knowledge we can directly understand e.g., the dependence of the Butcher-Oemler effect on such parameters as radius and magnitude cut.
%The strength of this method is apparent in that we have not only decreased the background subtraction problems, but also been able to separate different spectral populations of the cluster, making a more detailed study of the cluster population possible. 

We find a steep faint-end slope of the rest frame $B$--band cluster luminosity function, with $\alpha~\sim~-1.4$. This is consistent with the slope at faint magnitudes for both local and $0.1~\lsim~z~\lsim~0.3$ clusters. The steepness of the LF is found to be correlated to the radial distance from the cluster centre, with a more shallow slope near the core, compared to the outer parts of the cluster. This is closely related to the difference found in the radial distribution between bright and faint galaxies, where bright galaxies are more centrally concentrated. Most surveys reporting flatter slopes, $-1.2~\lsim~\alpha~\lsim~-1.0$, only cover the core region of the cluster, or do not probe the faint population beyond the upturn in the LF. 

When we divide the cluster galaxies spectroscopically into early-type and late-type galaxies, we find that these populations have very different LFs. Early-type galaxies, including ellipticals and red spirals, are best represented by a Gaussian. The late-type galaxies, including blue spirals, irregulars and dwarfs, are best fitted by a Schechter function with a steep faint-end slope, $\alpha~\sim~-1.7$.

We find that the 'plateau' region of the LF at intermediate magnitudes, found in most local and low--$z$ clusters, is absent in Cl1601+42, and we suggest that this can be due to a general brightening of the blue population of galaxies, caused by a higher star formation rate. If the progenitors of todays dwarf galaxies in local clusters have faded by $\sim$ 2--3 magnitudes since $z~\sim~0.5$, this class of galaxies could contribute to the steep slope seen in Cl1601+42.   
        
We find significant differences in the radial distributions when we divide our sample into bright and faint galaxies. Bright galaxies, which are predominantly early-types, are highly concentrated to the centre. This shows that the high density of early-type galaxies in the core of local clusters is established already at $z~\sim~0.5$. The faint galaxies have a more shallow radial decline at increasing radii. At the core the surface density flattens.

Bright, blue, late-type galaxies are rare in the cluster. This is consistent with the hierarchical scenario where in-falling spiral galaxies have their star formation quenched as they approach the centre of the cluster. Faint blue galaxies are more numerous, and are present across the whole cluster. There is, however, a decrease in the surface density at the core, which is consistent with a region close to the core depleted of faint, blue star forming galaxies. A possible mechanism could be the destruction of small galaxies through 'galaxy harassment'.

The fraction of blue galaxies in Cl1601+42 is $f_B~\sim~0.15$, which is higher than the local value, and qualitatively in agreement with the BO effect. The fraction is, however, lower than predicted from the average $f_B(z)$ relation found by BO84.
The blue fraction is found to depend on the distance to the core centre, in accordance with other studies. The increasing value of $f_B$ with increasing radius is consistent with the hierarchical scenario, where clusters are formed by accretion of blue field galaxies. $f_B$ is also found to increase with the limiting magnitude. This is a direct consequence of the different LF's for the early-type and late-type galaxies. The radial dependence and the shape of the LF, especially its faint population, are therefore connected.

Finally, we note that although this study goes deeper than most previous investigations at this redshift, the conclusions are only based on one cluster, and a large sample including clusters of varying richness and concentration is needed to access the robustness of our conclusions. We hope to address this in future studies.
\section*{Acknowledgements}
We are grateful to Claes-Ingvar Bj\"{o}rnsson and G\"{o}ran \"{O}stlin for comments on the paper, to the referee for several valuable comments and suggestions, and to the EIS project at ESO for making their data publicly available.
This work is supported by the Swedish Research Council and the Swedish Board for Space Sciences.
Nordic Optical Telescope is operated on the island of La Palma jointly by Denmark, Finland, Iceland, Norway, and Sweden, in the Spanish Observatorio del Roque de los Muchachos of the Instituto de Astrofisica de Canarias.

\label{lastpage}
\end{document}